\documentclass[preprint,12pt]{elsarticle}

\usepackage{array}
\usepackage{tabu}
\usepackage{caption}
\usepackage{subcaption}
\usepackage[T1]{fontenc}
\usepackage{url}
\usepackage{adjustbox}
\usepackage{multirow}
\usepackage{algorithmicx}
\usepackage[Algorithm,ruled]{algorithm}
\usepackage[noend]{algpseudocode}

\usepackage{amssymb}

\journal{}

\begin{document}

\begin{frontmatter}







\title{Exploiting Workload Cycles for Orchestration of Virtual Machine Live Migrations in Clouds}



\author[lab1]{Artur Baruchi\corref{cor1}}
\ead{artur.baruchi@gmail.com}

\author[lab1]{Edson T. Midorikawa}

\author[lab1]{Liria M. Sato}
\address[lab1]{University of Sao Paulo, Brazil}

\author[lab2]{Marco A. S. Netto}
\address[lab2]{IBM Research, Brazil}

\cortext[cor1]{Corresponding author}

\begin{abstract}
Virtual machine live migration in cloud environments aims at reducing energy costs and increasing resource utilization. However, its potential has not been fully explored because of simultaneous migrations that may cause user application performance degradation and network congestion. Research efforts on live migration orchestration policies still mostly rely on system level metrics. This work introduces an Application-aware Live Migration Architecture (ALMA) that selects suitable moments for migrations using application characterization data. This characterization consists in recognizing resource usage cycles via Fast Fourier Transform. From our experiments, live migration times were reduced by up to 74\% for benchmarks and by up to 67\% for real applications, when compared to migration policies with no application workload analysis. Network data transfer during the live migration was reduced by up to 62\%. 

\end{abstract}


\end{frontmatter}


\section{Introduction} \label{sec:Intro}

Through multiplexing techniques, virtualization allows for different operating systems and workloads to co-exist on the same hardware, without users' perception and interference among themselves. This feature is the core of cloud computing, which is a concept that dates back to the 1960s \cite{Parkhill:1966}.

Live migration also allows a Virtual Machine (VM) to be moved across physical hosts with minimal interruption. This feature brings several benefits to cloud providers, including policy creation for physical host maintenance and energy consumption reduction via server consolidation \cite{Srikantaiah:2008}. Another benefit is to distribute VM load among physical hosts to meet circumstantial computing demand \cite{Nuaimi:2012}.

Despite the resource usage optimization brought by server consolidation or load balancing policies, they still generate VM performance degradation \cite{Alan:2013}. This problem comes mainly from live migration algorithms, whose performance is sensitive to VM memory usage. These algorithms can generate large data traffic to migrate VMs across hosts, especially when several VMs are moved simultaneously.

To address this problem, our previous work \cite{Baruchi:2014} introduced an Application-aware Live Migration Architecture (ALMA) which determines, before hand, suitable moments to move VMs among physical hosts according to VM workloads. Our hypothesis is that, we can reduce live migration side effects, such as data traffic, by choosing the right moment to trigger the migration process. This hypothesis comes from two observations. The first is live migration algorithms are sensitive to VM memory usage and the second is several industry and scientific workloads follow a resource usage cyclic pattern. Examples of these cycle scenarios are: (i) a Web service with higher access during some periods of day or due to application characteristics that can present long periods of processor usage after I/O access and (ii) parallel applications with processes exchanging synchronization messages.

This work presents a detailed description of ALMA algorithms and the characterization process. Moreover, a new set of experiments to evaluate the effectiveness of using application characterization data to trigger migrations and a scalability analysis of our solution is presented. Therefore, we extended our work with the following contributions:
\begin{itemize}
    \item Characterization and workload cycle recognition using Fast Fourier Transform with benchmarks and real scientific applications ($\S$ \ref{sec:cycle});
    \item Live migration orchestration based on workload cycle recognition ($\S$ \ref{sec:orchestration});
    \item Evaluation of the architecture that implements the migration orchestration on a private cloud environment, including scalability analysis of the architecture with data from up to 1,000 VMs ($\S$ \ref{sec:evaluation}).
\end{itemize}
\section{Motivation and Problem Description}

The overhead of live migrations comes mainly from algorithms, like pre-copy \cite{Marvin:1985} and post-copy \cite{Hines:2009} and, at first, it is not the scope of server consolidation policies to deal with migration algorithm issues. As a result, server consolidation is barely used inside the cloud provider data centers \cite{Birke:2013}.

\begin{figure}[!h]
        \centering
        \begin{subfigure}[b]{0.49\textwidth}                        
                \includegraphics[width=\textwidth]{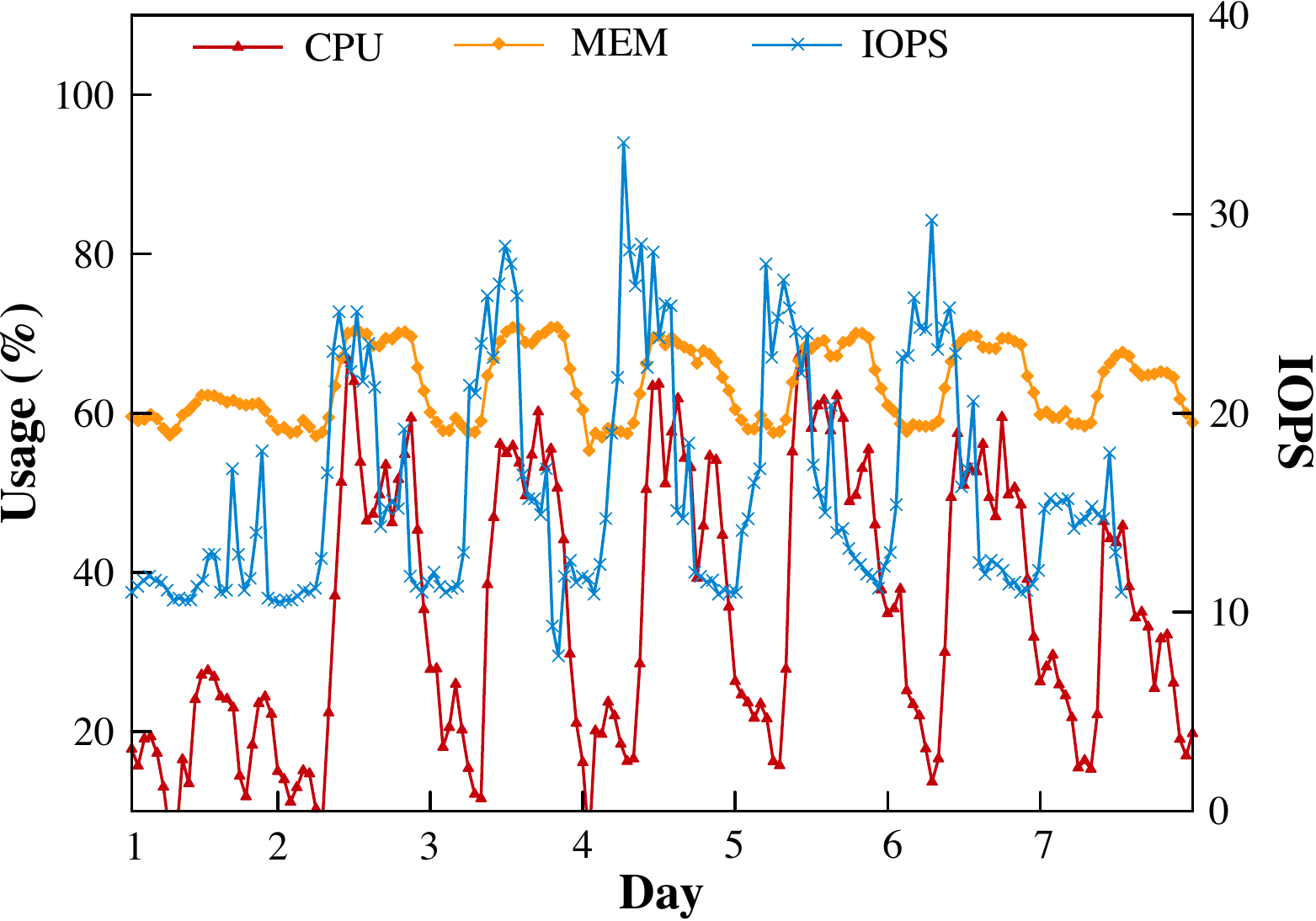}
                \caption{Data Base real workload.}\label{fig:ProducaoBCO}
        \end{subfigure}
        \begin{subfigure}[b]{0.49\textwidth}                         
                \includegraphics[width=\textwidth]{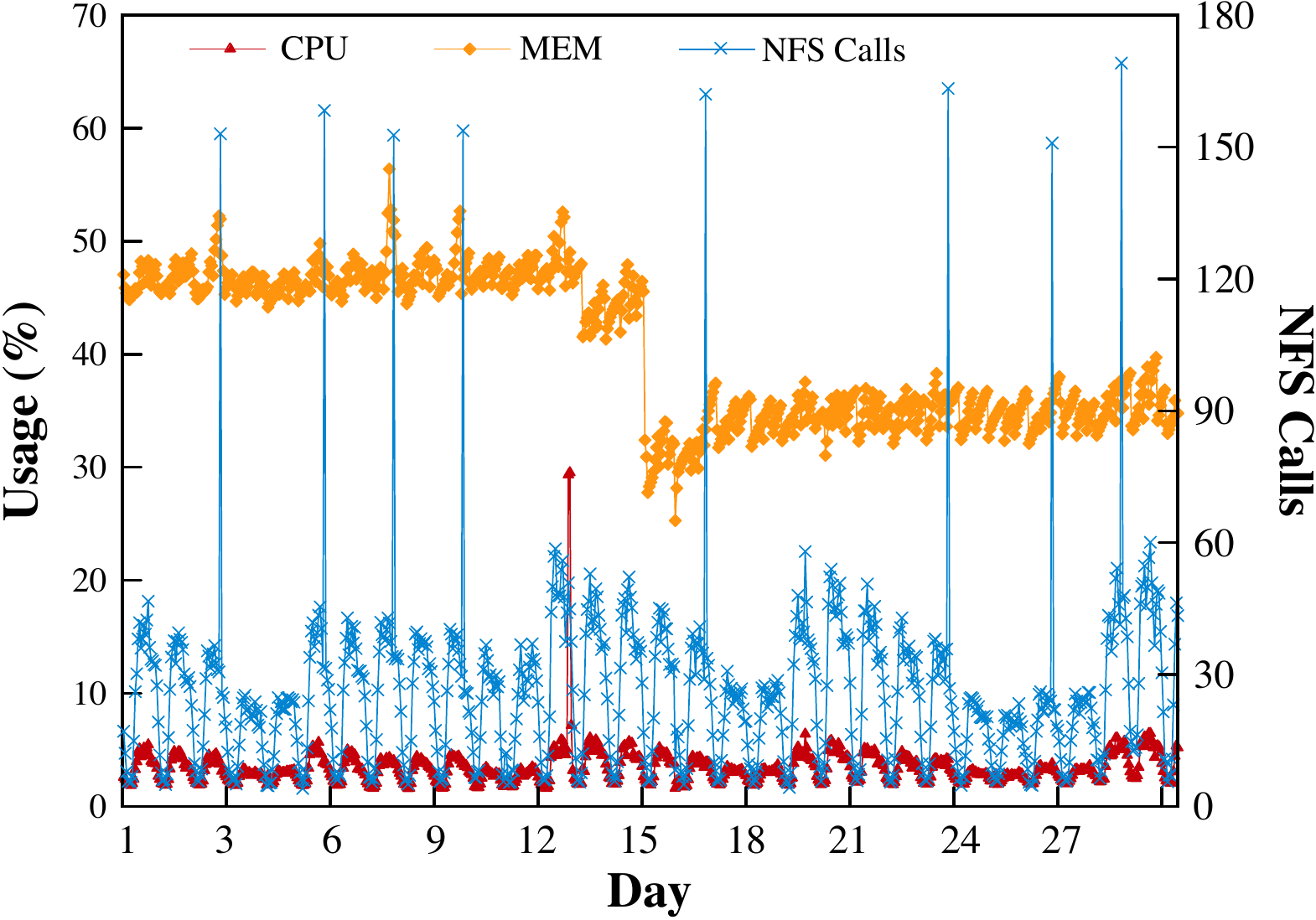}
                \caption{Web real workload.}\label{fig:producaoWEB}
        \end{subfigure}
        \caption{Real workloads with cyclic behaviors.}
        \label{fig:cargasDeProducao}
\end{figure}

The proposed architecture explores live migration algorithm characteristics and mediates between the consolidation policies and live migration algorithms, reducing the impact created by multiple concurrent migrations. Our architecture is based on the observation that several workloads have cyclic behaviors. Two examples of real workloads with cyclic behavior are presented in Figure \ref{fig:cargasDeProducao}, which illustrates the resource usage of a production data base from a telco company during a week and the resource usage of a pool of VMs of a big magazine publisher. We can see a cyclic behavior in both graphs and the best periods to perform live migrations is during the valleys of the graph, which consist of low resource usage.

\begin{figure}[!t]
        \centering
        \begin{subfigure}[b]{0.35\textwidth}                        
                \includegraphics[width=\textwidth]{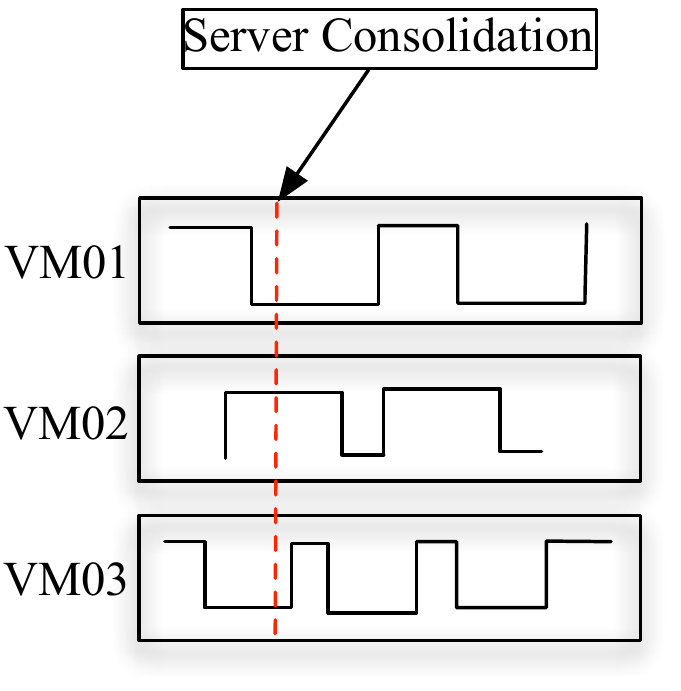}
                \caption{Without cyclic analysis.}\label{fig:WithoutCA}
        \end{subfigure}
        ~
        \begin{subfigure}[b]{0.35\textwidth}                         
                \includegraphics[width=\textwidth]{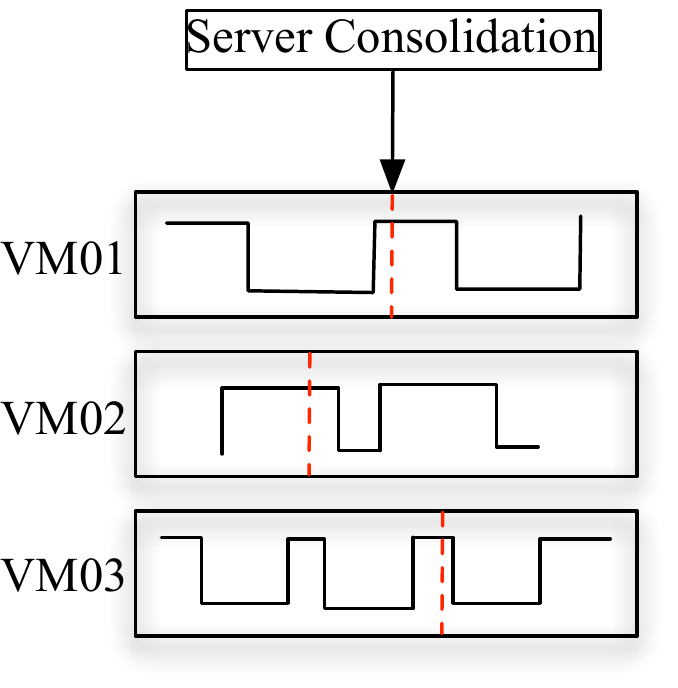}
                \caption{With cyclic analysis.}\label{fig:WithCA}
        \end{subfigure}
        \caption{Live migration orchestration using cyclic analysis: valleys represent moments where live migrations have potential to congest the network and peaks represent suitable moments for live migration.}
        \label{fig:LMOrchestration}
\end{figure}

Figure \ref{fig:LMOrchestration} shows two scenarios for the timing of triggering migrations: one in which the consolidation triggers live migration just after the definition of the new physical hosts (without cyclic analysis), and the other scenario, where VM live migrations are orchestrated according to the their workload (with cyclic analysis). In the first scenario, live migrations produce more network traffic, since two (VM01 and VM03) out of three VMs are migrated during a not suitable moment. Moreover, if VM03 live migration were postponed, its workload would be in a suitable moment, avoiding network congestion---which is the second scenario.

The main goal of the cyclic analysis is to identify and extract the workload's execution pattern and postpone live migrations with potential to harm the network. In Figure \ref{fig:WithCA}, all live migrations were postponed to a suitable moment to reduce live migration time and network congestion. With our architecture, algorithms for server consolidation and live migration (pre-copy and post-copy) are not modified. Our solution intercepts all pending migrations, and orchestrates them according to the workload cycles. In practice, there should be modification only in the consolidation strategy APIs for concurrent live migrations.

The problems tackled in this paper are therefore: (i) how to detect application resource consumption cycles and (ii) how to use this information to know suitable moments for moving VMs across physical hosts in order to minimize network traffic and migration times.

\section{Background} \label{sec:Sec01}

In this section we discuss concepts to understand how our architecture can reduce live migration overhead. These concepts cover pre-copy live migration algorithm and the basics of server consolidation.

\subsection{Live Migration}

The encapsulation of the operating system execution environment offered by virtualization is fundamental to live migrations \cite{Smith:2005}. This feature allows for: (1) at any time, a VM to be frozen, (2) all necessary information from restarting the execution to be stored in a file and (3) using this information to restart a VM on any physical host from the breakpoint.

Live migration algorithms can be classified according to the moment in which VM's memory content is copied to its replica on the destination host. The main migration algorithms are: \textit{pre-copy} and \textit{post-copy}. The pre-copy algorithm copies VM memory before it starts its execution in the destination host, whereas the post-copy copies VM memory after it starts its execution.

Research studies use four metrics to compare these algorithms---two are related to performance evaluation and two refer to VM workload overhead caused by live migration:

\begin{itemize}
    \item {\bf Live migration total time:} time interval between the start of the migration process and the VM execution beginning in destination host;
    \item {\bf Downtime}: time interval in which VM is not running nor available to the user;

    \item {\bf Execution time:} execution time of an application, including a possible migration time;
    
    \item {\bf Throughput:} amount of data processed in a time interval with and without live migrations.
\end{itemize}

\subsection{Pre-Copy}\label{subsec:pre-copy}

As pre-copy moves VMs only when their memory is already copied, this algorithm is more robust and widely used in commercial Virtual Machine Monitor (VMM) compared to post-copy---we used this algorithm in this paper. The memory is copied between hosts in several iterations and can be split into five stages \cite{Kikuchi:2012}:

\begin{enumerate}
    \item {\bf Resource reservation:} it checks whether destination host has available resources to the VM;
    \item {\bf Iterative copy:} the VM's memory is entirely copied to destination host in the first iteration. In next iterations only the memory changed in last iteration is copied (dirty pages);
    \item {\bf Stop and copy:} the VM is suspended in source host and the last copy is performed;
    \item {\bf Shutdown:} the VM is stopped in source host and all resources allocated to the VM are released;
    \item {\bf Activation:} the VM is activated in destination host.
\end{enumerate}    
    
One of the main problems of this algorithm is the possibility of unlimited cycles of memory copy. To avoid this issue, some VMMs impose conditions to stop the copy iterations. Taking Xen VMM \cite{Barham:2003} as example, the stop conditions are: (i) less than 50 pages marked as dirty since the last iteration; (ii) maximum of 29 iterations; and (iii) amount of data transferred greater than three times of memory assigned to the VM.

Furthermore, this algorithm is sensitive to VM dirty page rate and network throughput. Some studies, such as the one from Strunk \cite{Strunk:2012}, formalized the dirty page rate and network throughput dependencies. The author defines an upper and lower limit of the pre-copy algorithm that are presented in Inequalities \ref{eq:limiteTempoMigracao} and \ref{eq:limiteDowntime} containing limits to migration and downtime time:

\begin{equation} \label{eq:limiteTempoMigracao}
\frac{V_{mem}}{B} \leq T_{mig} \leq \frac{(M+1)*V_{mem}}{B}
\end{equation}
\begin{equation} \label{eq:limiteDowntime}
0 \leq T_{down} \leq \frac{(M+1)*V_{mem}}{B}
\end{equation}\\
Where:\\
$V_{mem}$: amount of memory assigned to VM to be moved;\\
B: network throughput available;\\
$T_{mig}$: live migration duration;\\
$T_{down}$: downtime duration;\\
M: number of times allowed to copy the entire memory in iteration phase.\\

The lower limits of Inequalities \ref{eq:limiteTempoMigracao} and \ref{eq:limiteDowntime} refer to an idle VM. In this situation, the migration duration is limited only by the network throughput between the two hosts and the downtime duration due to the low dirty page rate. However, the upper limit is related to a high dirty page rate, leading to the occurrence of several copy iterations. The worst case happens when the dirty page rate is higher than network throughput.

There are other factors that influence live migration duration, as observed by Xu et al. \cite{Xu:2013}, such as the number of concurrent migrations. However, the dominant factor to live migration performance is the dirty page rate.

\subsection{Server Consolidation}
Server consolidation policies consist in choosing VMs, according to a given criterion, and concentrating them into a few physical hosts. Server consolidation is fundamental to help cloud providers reduce energy consumption. The main problem of consolidation policies is to find an optimal combination of VM placement in physical hosts. Another dilemma is how to avoid concurrent migrations to accomplish an objective \cite{Ferreto:2011}.

The most common policies of consolidation are based on heuristics \cite{Feller:2012} or linear programming \cite{Ferreto:2011}, where the former is more explored by researchers due to scalability issues. Consolidation based on heuristics are more flexible and the final solution (usually suboptimal) is obtained faster. Implementations based on linear programming are more efficient when dealing with several restrictions, such as Service Level Agreements (SLAs) and maximum number of concurrent live migrations. 
\section{Workload Characterization and Cycle Recognition} \label{sec:cycle}

Workload characterization is the strategy used to collect information about what resources and their consumption by applications under analysis. An application can use several computing resources at the same time, but it is likely that a specific resource is being used more than others. This behavior can be static, meaning that an application can use a given  resource more from the beginning to the end or dynamic, when during the application execution, it can use different resources and at various utilization levels. In this work, the workload characterization is defined in time unit to perceive fluctuations in resource usage during the application run time. We characterize the workload every fifteen seconds.

Server consolidation relies on workload characterization to avoid the placement of VMs competing for the same resources in the same physical hosts. There are challenges in cloud workload characterization due to the features of this paradigm, such as dynamic resource usage \cite{Armbrust:2009} and multi-tenancy. These factors can produce ambiguous signals to the workload classifier and generate wrong results.

Another common challenge in workload characterization is the data interpretation and gathering from VMs. Virtual Machine Monitors (VMMs) add a second level of indirection in order to isolate VMs located in the same host. This extra level, known as Semantic Gap \cite{Garfinkel:2005}, imposes several challenges to interpret and gather performance metrics. 

The majority of traditional workload characterization strategies are computationally expensive and prohibitive to be implemented in a cloud data center running thousands of VMs simultaneously. Recent research findings specialized in VM characterization are difficult to implement in the cloud, because such strategies are mostly based on specific VMM metrics \cite{Ahmad:2007} or in VMM's source code modification \cite{Du:2011}.

A load index is a metric that aims at quantifying the system load in a given moment or during a time interval. In this work we used load indexes related to processor and memory resources, which are enough to identify other types of load, such as I/O.

\subsection{Naive Bayes Classifier}

The Naive Bayes (NB) classifier is based on Bayes's theorem, which is broadly used in the probability field. The Naive term is due to the assumption the events' probability are independent of one another.

The aim of Bayesian classifiers is to estimate the most probable class of a set of characteristics using probabilities known {\it a priori}, which are computed used a training data set. The Bayes's theorem requires at least three terms---one conditional probability and two unconditional probabilities---to compute a third conditional probability \cite{Russell:2009}.

The quantitative results of the NB classifier is one of its main features. When submitting data to the classification, the NB classifier returns the most likely class of that data and their probability, allowing the implementation of optimization strategies.

\subsection{Cyclic Analysis}
Once the workload characterization has been completed as LM (live migration) or NLM (no live migration), cyclic patterns can be extracted, if any, from the characterization data collected over time. Given that possible classes are only LM or NLM, workload cycles can be identified and decomposed.

The extraction process is made by Fast Fourier Transform (FFT) \cite{Bergland:1968}. FFT has $O(n\log{}n)$ complexity where $n$ is the number of samples used to compute the cycle. FFT allows to convert time (or space) in frequency.

\begin{figure}[!h]
    \centering
    \includegraphics[width=0.8\textwidth]{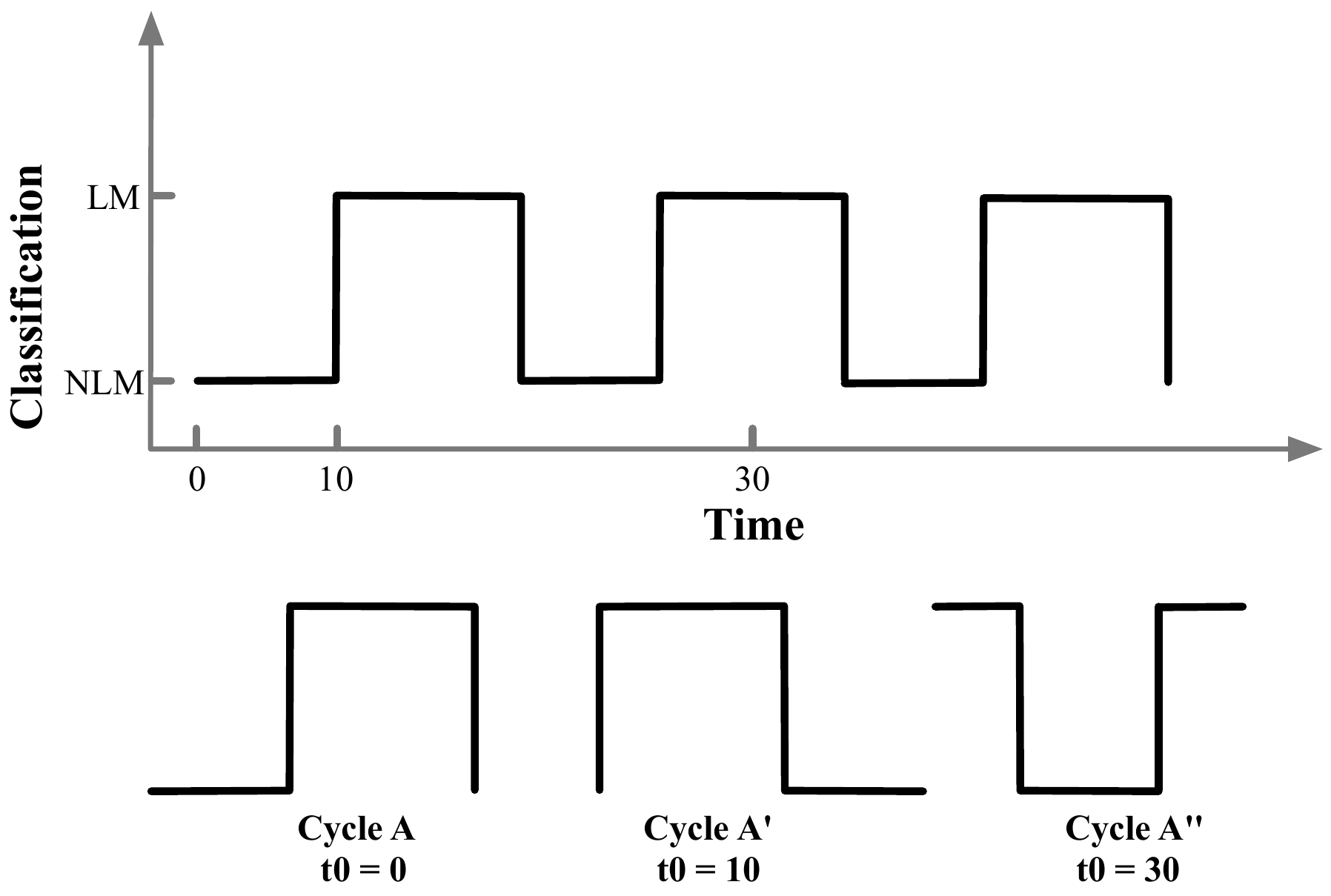}
    \caption {Shapes for the same cycle.}\label{fig:SameCycleComp}
\end{figure}

In this work we define a cycle as a recurrent pattern of a workload, which can be composed of several moments, suitable or non-suitable to live migration. Additionally, in a same cycle, there can be several compositions, as presented in Figure \ref{fig:SameCycleComp}. The cycle shape depends on the initial instant $t_{0}$. If $t_{0}$ is in 0, the cycle's shape will be like A, if $t_{0}$ is in 10 or 30, the cycle's shape will be like A' and A'', respectively.

\begin{figure}[!h]
    \centering
    \includegraphics[width=0.8\textwidth]{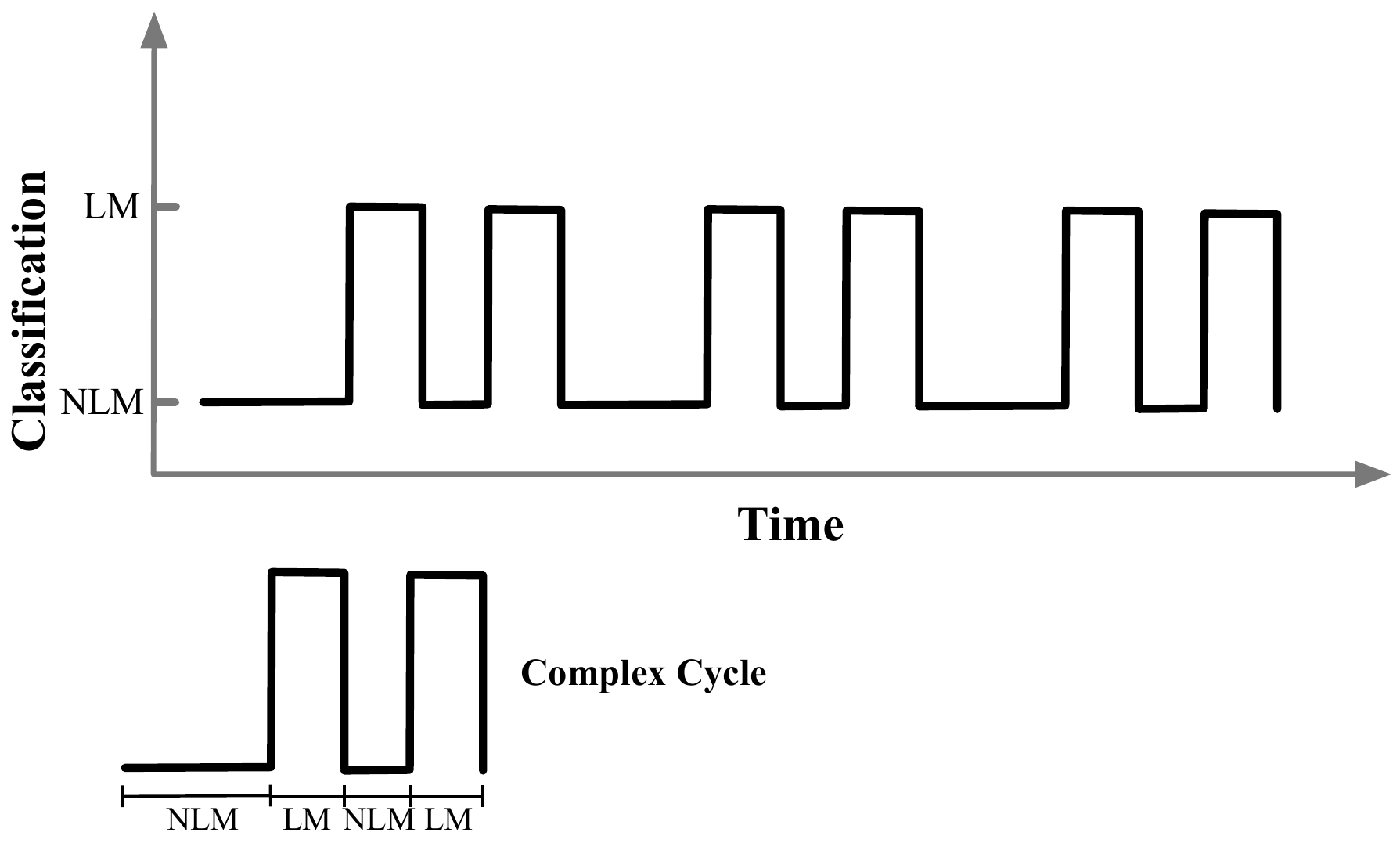}
    \caption {A complex cycle.}\label{fig:ComplexCycle}
\end{figure}

A cycle can be simple or complex. A simple cycle is composed of up to three interleaved intervals, that is, there is a single occurrence of one type of workload (LM or NLM). In Figure \ref{fig:SameCycleComp} the cycle is simple, even in shape A'', where we observe three interleaved intervals (LM, NLM and LM). In Figure \ref{fig:ComplexCycle} is presented a complex cycle example, which contains two intervals of NLM and LM. FFT can identify both types of cycles. 

\section{ALMA - Application Aware Live Migration Architecture} \label{sec:orchestration}

Cyclic analysis, based on workload characterization, can be used to define suitable moments to trigger live migrations. To this end, we propose an architecture, called Application-aware Live Migration Architecture (ALMA) \cite{Baruchi:2014}, which intermediates all live migration requests from massive migration strategies and the VM monitor.

\subsection{Architecture Overview}

Efforts to solve macro problems in a data center, such as energy consumption, computational waste and live migration algorithm optimizations do not address problems related to pre-copy and post-copy algorithm's limitations. Our architecture avoids live migration drawbacks by choosing suitable moments to trigger this operation, which benefits strategies broadly used to solve macro problems.

\begin{figure}[!h]
    \centering
    \begin{subfigure}[b]{0.48\textwidth}                        
        \includegraphics[width=\textwidth]{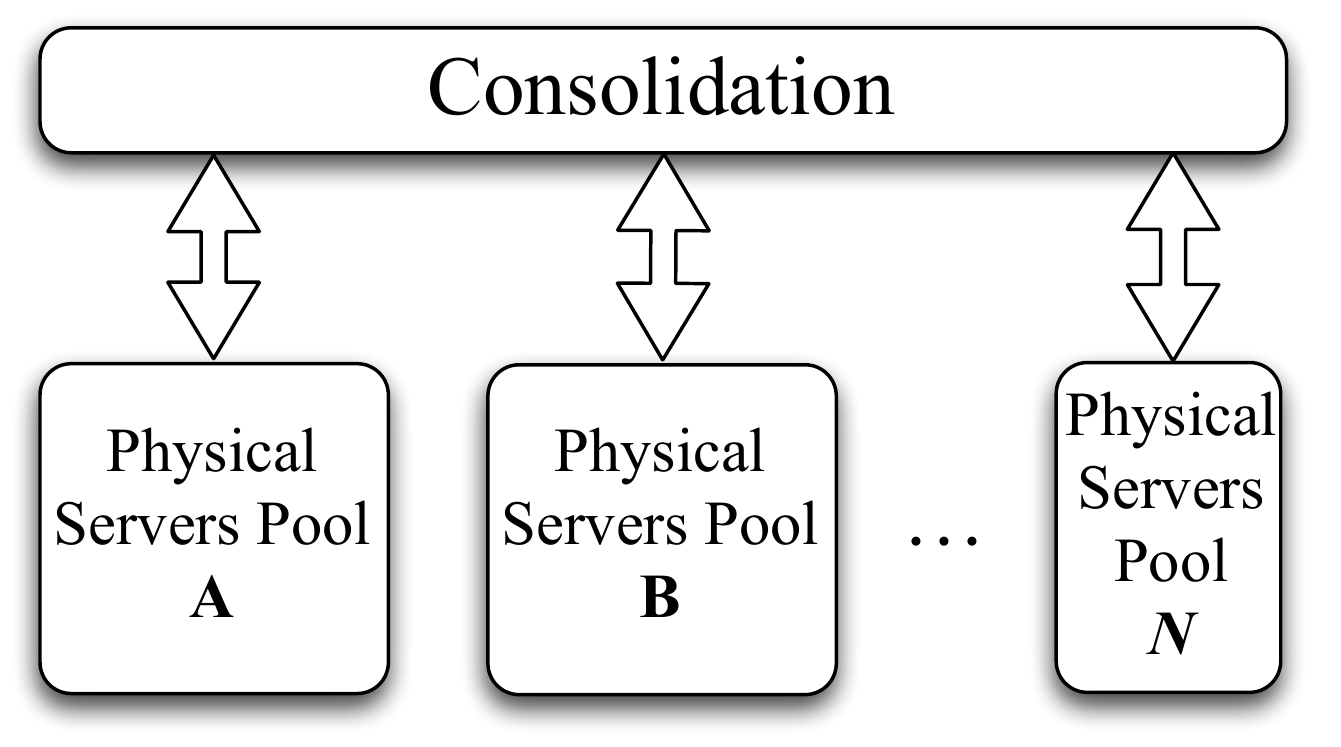}
        \caption{No live migration control.}\label{fig:NoLMControl}
    \end{subfigure}
    ~
    \begin{subfigure}[b]{0.48\textwidth}                         
        \includegraphics[width=\textwidth]{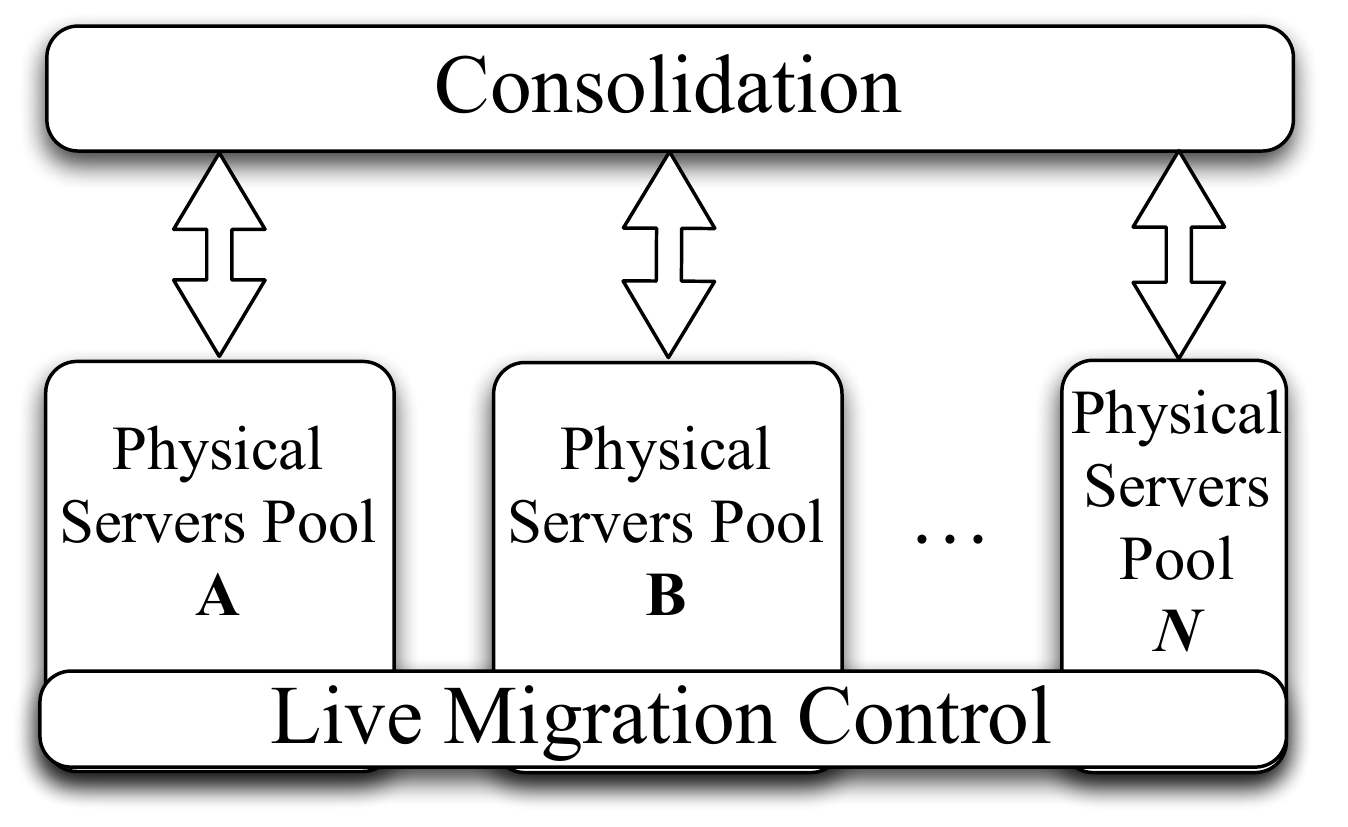}
        \caption{Traditional live migration control.}\label{fig:ControleLMClassico}
    \end{subfigure}
    \newline
    \begin{subfigure}[b]{0.48\textwidth}                         
        \includegraphics[width=\textwidth]{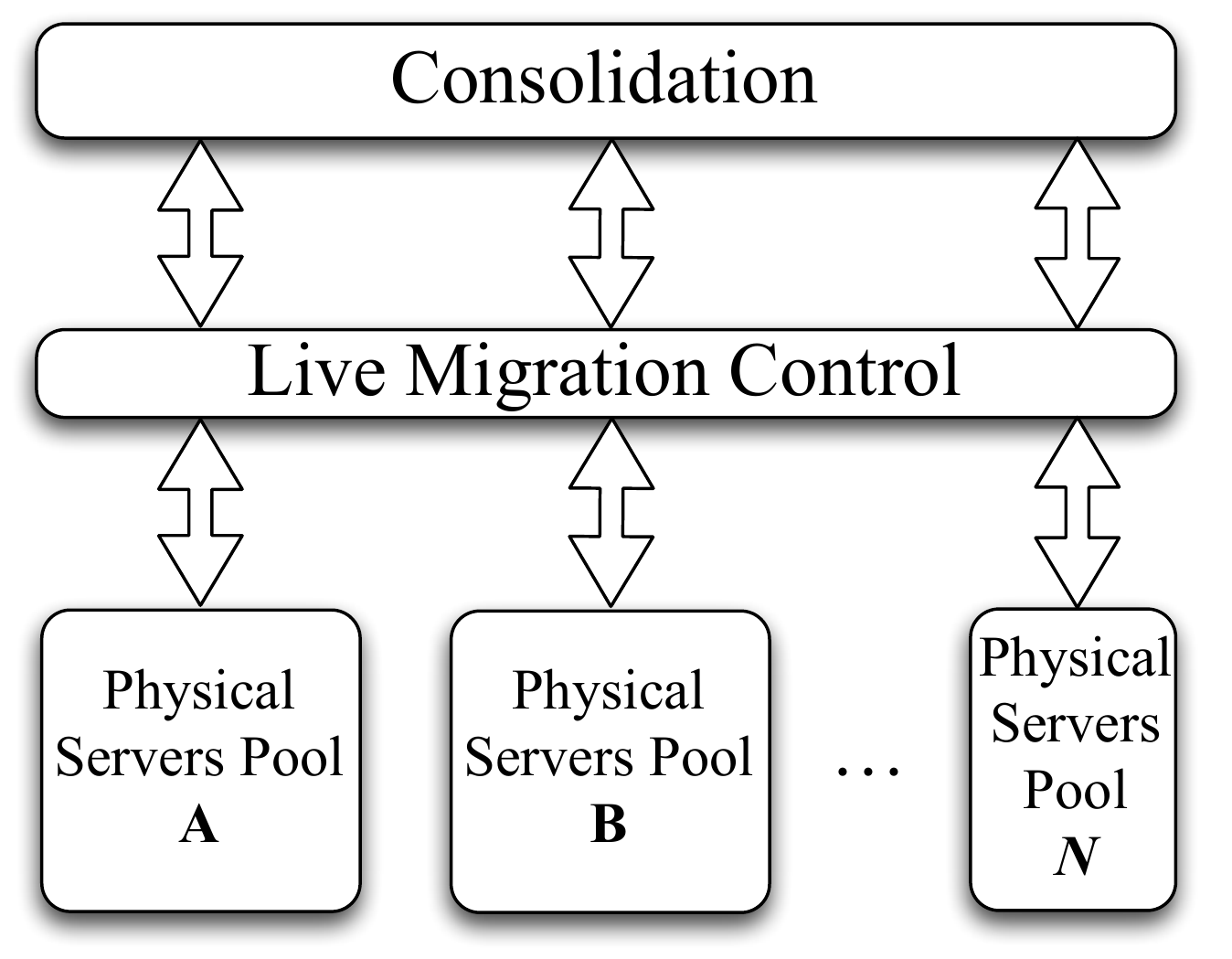}
        \caption{ALMA Proposal.}\label{fig:ALMAProposal}
    \end{subfigure}              
    \caption{Architectures for virtual machine live migration.}
    \label{fig:Controls}
\end{figure}

Figure \ref{fig:Controls} illustrates ALMA and the other two most common architectures for live migration. Figure \ref{fig:NoLMControl} presents an architecture with no live migration control---once a VM needs to be moved across hosts, the architecture does so without any concern about network traffic and other ongoing live migrations \cite{Seki:2012}. The architecture presented in Figure \ref{fig:ControleLMClassico} implements a control over the live migrations. However, this control comes from the VM monitor and it orchestrates ongoing live migrations. The orchestration in this architecture considers only one or two metrics, such as available network bandwidth.    

Our architecture, presented in Figure \ref{fig:ALMAProposal}, has a different approach. When the live migration plan is created, the architecture receives all live migration submissions and orchestrates the migrations. The main component of ALMA is the Live Migration Control Module (LMCM), which is responsible for migration scheduling based on the VM workload. It can postpone, run immediately or cancel a live migration. The postponement can occur when the VM workload is under a not suitable moment to trigger live migration and then it has to wait for a better moment. However, if the workload is suitable to be moved, ALMA  triggers the migration immediately. If the workload is almost at the end and the live migration cost is higher than keeping the VM running on the current physical server, ALMA can cancel the live migration request.

LMCM accepts parameters from cloud service provider to impose limits, such as the maximum time a VM can wait to be migrated. This could avoid long waiting time due to long cycle periods. On the customer side, there are some parameters that could be implemented, for example the expected time to finish a given workload and with this information ALMA could avoid migrations that harm customer-defined thresholds.

\subsection{Algorithms}   

ALMA implementation is based on two algorithms. The first one finds and extracts cyclic pattern from a workload, while the second computes the waiting time for a suitable moment for live migration. These algorithms are complementary to one another, since the output from one is used by the other and could be implemented together. However, for a better understanding of the strategy, they will be described separately.

The first algorithm uses the workload classification collected for a given time interval, which is sorted chronologically. Each array position has a VM characterization for the period of the data collected.  Moments are represented by the array index and the Fast Fourier Transform computes the cycle size (line 2) based on this array. Thereafter, all analyses occur in the interval of the array that represents the cycle size.

\begin{algorithm}[!h]
    \begin{algorithmic}[1]
        \Require An array $C$ with VM workload classification data for a given time interval. The array should be chronologically ordered. 
        \Function{Decomposition}{$C$}
        \State $CycleSize \gets FFT(C)$ \Comment{Fast Fourier Transform.}
        \State $LMCount \gets 1$
        \State $NLMCount \gets 1$
        \For{i}{1}{$CycleSize$} 
        \If {$C[i] == LM$}
        \State $ArrayLM[LMCount] \gets i$
        \State $LMCount \gets LMCount + 1$
        \Else
        \State $ArrayNLM[NLMCount] \gets i$
        \State $NLMCount \gets NLMCount + 1$
        \EndIf
        \EndFor\\
        \Return {$ArrayNLM, ArrayLM$}
        \EndFunction
    \end{algorithmic}
    \caption{Cycle decomposition in two arrays.}\label{alg:decomposicaoVetor}
\end{algorithm}

The part of the array representing an entire cycle is then split into two smaller arrays (line 5 to 13). One array stores only suitable moments for live migration (ArrayLM) and the other stores moments not favorable for live migration  (ArrayNLM).

The second algorithm aims to find the instant in which the workload is inside a cycle. We use two known variables: (1) the cycle length time, which is computed in the first algorithm and (2) the workload execution time, which is the elapsed time of the workload execution. The relative time ($M_{relative}$) can be computed as the module between elapse time of the workload ($M_{current}$) and the cycle length time ($CycleSize$, line 2 of Algorithm \ref{alg:instanteLM}).

After the computation of $M_{relative}$, the next step is to find in which array it is placed (ArrayLM or ArrayNLM). If the relative moment is placed in ArrayNLM (line 3), we need to find out when the workload will be in suitable moment to be migrated (ArrayLM). In order to compute the remaining time of this period, we look for the first instant longer than the relative moment in ArrayLM (NextLM, in line 4). The difference between both instants (NextLM and $M_{relative}$) is the remaining time ($RemainTime$) to the workload be feasible to be migrated.

\begin{algorithm}[!h]
    \begin{algorithmic}[1]
        \Require Cycle size and current moment.
        \Function{Postpone}{$CycleSize, M_{current}$}
        \State $M_{relative} \gets M_{current}$ \% $CycleSize$
        \If {$find (M_{relative}, ArrayNLM)$}
        \State $NextLM \gets findGreater (M_{relative}, ArrayLM)$
        \State $RemainTime \gets NextLM - M_{relative}$
        \Else
        \State $RemainTime \gets 0$
        \EndIf\\
        \Return {$RemainTime$}
        \EndFunction
    \end{algorithmic}
    \caption{Identification of live migration moment.}\label{alg:instanteLM}
\end{algorithm}
\section{Evaluation} \label{sec:evaluation}

Our previous results showed substantial reduction in network data traffic and live migration time when using workload cycle recognition for live migration \cite{Baruchi:2014,Baruchi:2015}. Here we present a more detailed evaluation with more VMs, additional applications, and new insights and discussions.

Metrics evaluated in this work are:
\begin{itemize}
    \item {\bf Total migration time (secs):} time between the start of a migration submission and the moment the VM is released from the source host;
    \item {\bf Downtime duration (secs):} time in which the migrated VM is unreachable from network. Data is collected using ICMP;
    \item {\bf Network data transfer (MB):} amount of data transferred in the network during the live migration;
    \item {\bf Cycle accuracy identification:} used to show moments where migration actually occurred---not when they are requested.
\end{itemize}

We organized our experiments in three parts: (i) workload characterization and cycle recognition analysis; (ii) orchestration analysis with benchmarks and real scientific applications; and (iii) scalability tests to handle data from hundreds of VMs.

\subsection{Experiment Setup}

We setup a private cloud with five physical hosts, one Network Attached Storage (NAS) and ten VMs equally distributed among physical hosts. We used VMs with three computational resource configurations (Table \ref{tab:VMTestBedConf}). 

\begin{table}[!h]
\centering
\caption {Virtual Machine configurations used in testbed.}
\begin{adjustbox}{width=0.70\textwidth}
\begin{tabu}{cccc}
\hline
\hline
\textbf{Configuration} & \textbf{VCPUS} & \textbf{\begin{tabular}[c]{@{}c@{}}Memory\\ (MB)\end{tabular}} & \textbf{\begin{tabular}[c]{@{}c@{}}Virtual Machine\\ ({\it hostname})\end{tabular}} \\ \hline
Small               & 1              & 768                                                             & \begin{tabular}[c]{@{}c@{}}vm02\_A vm03\_A\\ vm01\_B vm02\_B\end{tabular}     \\ 
Medium                 & 2              & 1024                                                            & \begin{tabular}[c]{@{}c@{}}vm01\_A vm01\_C\\ vm01\_D vm02\_D\end{tabular}     \\ 
Large                & 2              & 2048                                                            & vm03\_B vm02\_C                                                               \\ \hline
\end{tabu}
\end{adjustbox}

\label{tab:VMTestBedConf}
\end{table}

\begin{figure}[!t]
        \centering
        \begin{subfigure}[b]{0.48\textwidth}                        
                \includegraphics[width=\textwidth]{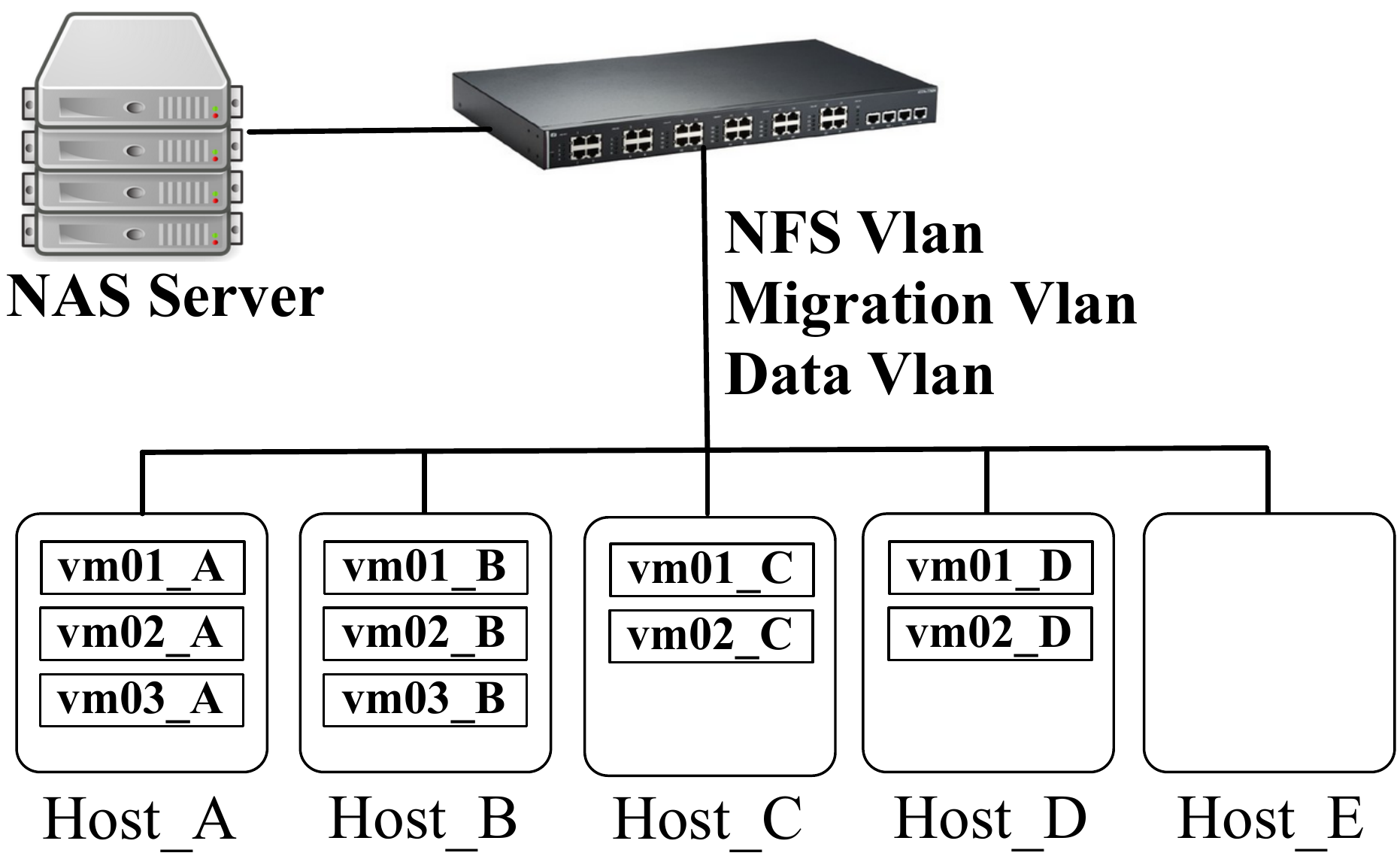}
                \caption{Before consolidation.}\label{fig:BeforCons}
        \end{subfigure}
        ~
        \begin{subfigure}[b]{0.48\textwidth}                         
                \includegraphics[width=\textwidth]{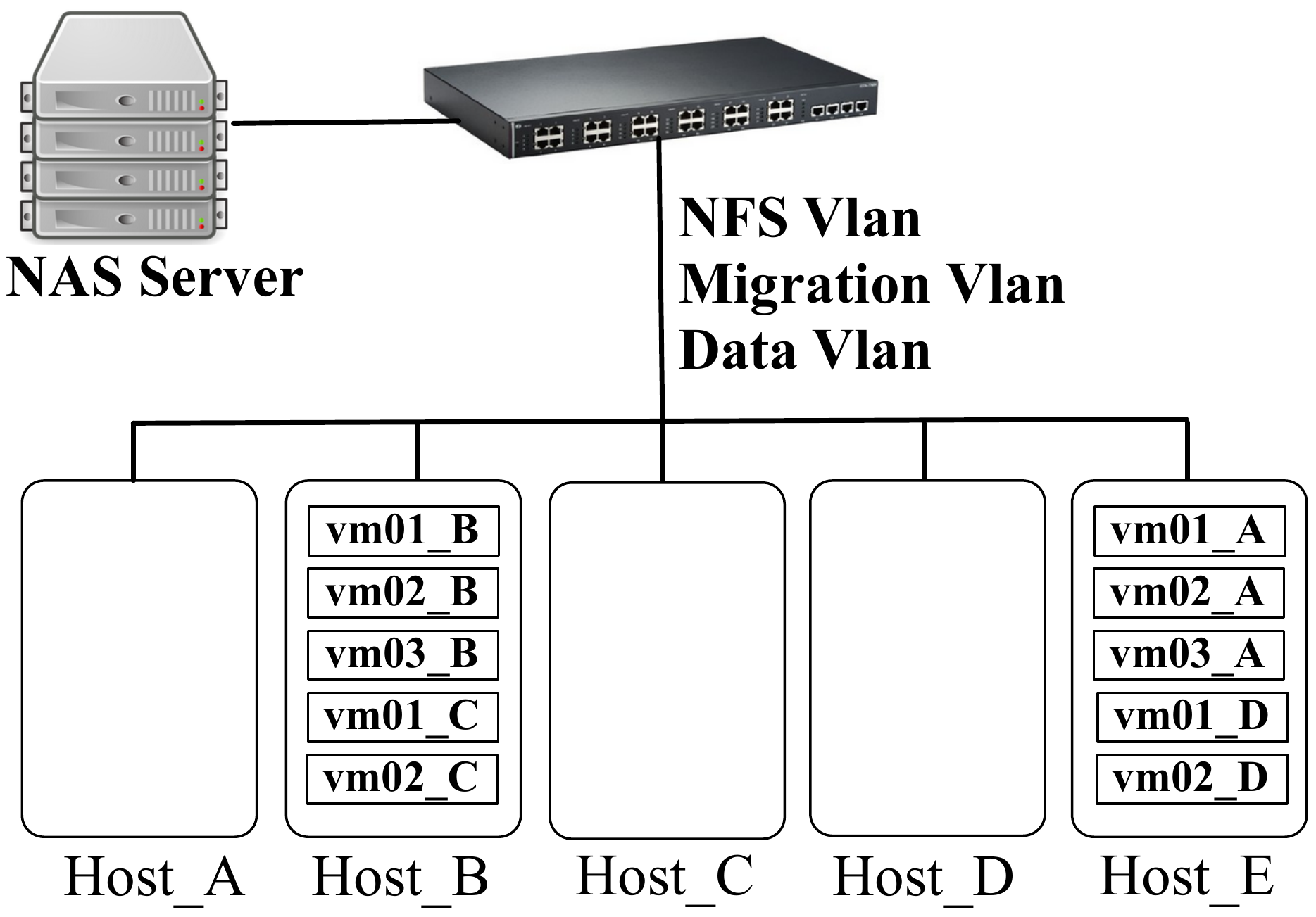}
                \caption{After consolidation.}\label{fig:AfterCons}
        \end{subfigure}
        \caption{Testbed topology.}
        \label{fig:Topology}
\end{figure}

VMs were initially placed in four physical hosts and during the workload execution, they were consolidated into two physical hosts (Figure \ref{fig:Topology}). The consolidation moments were randomly chosen to explore various points in time, having preferences for points where all machines were running workloads in order to stress the consolidation policies.

\begin{table}[!ht]
\centering
\caption{Benchmarks used in testbeds.}
\begin{adjustbox}{width=1\textwidth}
\label{tab:BenchmarksAppl}
\begin{tabu}{clc}
\hline\hline
{\bf \begin{tabular}[c]{@{}c@{}}Benchmark /\\ Application\end{tabular}} & \multicolumn{1}{c}{{\bf Description}}                                                                                                                                                                                                        & {\bf Experiment}                                                                                           \\ \hline 
\begin{tabular}[c]{@{}c@{}}TeraSort \\ (MapReduce)\\ \end{tabular}    & Sort algorithm which uses MapReduce paradigm \cite{Dean:2008}.\vspace{4mm}                                                                                                                                                                                              & \begin{tabular}[c]{@{}c@{}} Scientific \\ Application Evaluation\end{tabular}                         \\ 
\begin{tabular}[c]{@{}c@{}}BRAMS \\ \end{tabular}                     & Brazilian atmospheric model used for weather forecast \cite{Freitas:2009}.\vspace{4mm}                                                                                                                                                                               & \begin{tabular}[c]{@{}c@{}} Scientific \\ Application Evaluation\end{tabular}                         \\ 
\begin{tabular}[c]{@{}c@{}}OpenModeller \\ \end{tabular}              & Scientific application used to mode specimen distribution \cite{Souza:2011}.\vspace{4mm}                                                                                                                                                                       & \begin{tabular}[c]{@{}c@{}} Scientific \\ Application and \\ Characterization Evaluation\end{tabular} \\ 
\begin{tabular}[c]{@{}c@{}}SPEC CPU 2k \\ \end{tabular}               & \begin{tabular}[c]{@{}l@{}}It is a broadly used benchmark to compare computational systems.\\ It has several subprograms which stress the processor \cite{Sair:2000}.\vspace{4mm}\end{tabular}                                                                             & \begin{tabular}[c]{@{}c@{}} Benchmark and\\ Characterization Evaluation\end{tabular}                  \\ 
\begin{tabular}[c]{@{}c@{}}BT \\ \end{tabular}                        & \begin{tabular}[c]{@{}l@{}}Part of NASA Parallel Benchmark. This program has a memory\\ footprint of 650 MB with high rate of dirty page \cite{Bailey:1991}.\vspace{4mm}\end{tabular}                                                                                 & \begin{tabular}[c]{@{}c@{}} Benchmark \\ Evaluation\end{tabular}                  \\ 
\begin{tabular}[c]{@{}c@{}}IOZONE\\ \end{tabular}                    & \begin{tabular}[c]{@{}l@{}}Benchmark with high usage of I/O subsystem. To avoid cache usage\\ effect, files are larger than available memory \cite{Tarasov:2011}.\vspace{4mm}\end{tabular}                                                                         & \begin{tabular}[c]{@{}c@{}} Benchmark\\ Evaluation\end{tabular}                                       \\ 
sleep\footnotemark                                                                   & Linux command used to delay processing for a given time. \vspace{3mm}                                                                                                                                                                                      & \begin{tabular}[c]{@{}c@{}} Benchmark\\ Evaluation\end{tabular}                                       \\ 
\begin{tabular}[c]{@{}c@{}}LAME\\ \end{tabular}                      & \begin{tabular}[c]{@{}l@{}}LAME is an MP3 codifier used as benchmark \cite{Johnston:1992}.\\ We used input file of 2.3 GB.\vspace{4mm}\end{tabular} & \begin{tabular}[c]{@{}c@{}}Characterization\\ Evaluation\end{tabular}                                     \\ \hline
\end{tabu}
\end{adjustbox}
\end{table}
\footnotetext{Available at: \url{http://man7.org/linux/man-pages/man3/sleep.3.html}}

Benchmarks and applications (Table \ref{tab:BenchmarksAppl}) were used to evaluate the characterization strategy and ALMA. OpenModeller, LAME, and SPEC were used with different VM configurations for workload characterization. The ALMA evaluation consists of two experimental scenarios in order to create a controlled scenario. In first scenario we created artificial cycles running benchmarks with a specific behavior in a given order. The evaluation contains the SPEC benchmark as CPU intensive workload, BT as Memory intensive workload, IOZone as I/O intensive workload and, \texttt{sleep} command to simulate IDLE periods. The artificial cycles and VMs used are described in Table \ref{tab:ArtificialCycles}. The second scenario contains BRAMS, OpenModeller, and TeraSort where the former two are scientific applications and the latter represents a typical data-intensive cloud workload, all running simultaneously in different VMs.

\begin{table}[h]
\centering
\caption{Artificial cycles used to evaluate ALMA.}
\label{tab:ArtificialCycles}
\begin{adjustbox}{width=0.5\textwidth}
\begin{tabular}{cc}
\hline\hline
{\bf \begin{tabular}[c]{@{}c@{}}Virtual Machine\end{tabular}} & {\bf Artificial Cycles}                                                                       \\ \hline
vm03\_A                                                         & \begin{tabular}[c]{@{}c@{}}I/O CPU CPU I/O CPU\\ CPU I/O CPU CPU\vspace{4mm}\end{tabular}                    \\ 
vm02\_C                                                         & \begin{tabular}[c]{@{}c@{}}MEM IDLE CPU MEM IDLE CPU\\ MEM IDLE CPU MEM IDLE CPU\vspace{4mm}\end{tabular} \\ 
vm02\_A                                                         & \begin{tabular}[c]{@{}c@{}}MEM CPU CPU MEM CPU CPU\\ MEM CPU CPU MEM CPU CPU\vspace{4mm}\end{tabular}     \\ 
vm01\_C                                                         & MEM IDLE CPU MEM IDLE CPU                                                                     \\ \hline
\end{tabular}
\end{adjustbox}
\end{table}

\subsection{Workload Classification and Cycle Recognition}

The evaluation of the NB classifier was based on two benchmarks and one scientific application. Benchmarks behavior is more constant during the execution and, due to this characteristic, we can verify the NB classifier precision. On the other hand, the scientific application presents several oscillations during its execution and enables the analysis of the classifier sensibility to peaks of usage and workload changes.

A new subclass of four VM configurations, summarized in Table \ref{tab:configuracaoMV}, was used for this experiment. For each configuration, benchmarks (SPEC and LAME) and an application (OpenModeller) were run ten times and during the test load indexes were collected. VMs were installed in a hardware consisting of a 2.66 MHz Intel Core 2 Quad processor, 2 GB of memory, and a 5400 RPM hard disk with nominal throughput of 3 GB/second. As software configuration, the VMM used was Xen 4.1.3 and OpenSuse 12.1 running Kernel 3.1.10 as host OS. All VM images use CentOS 5.9 running Kernel 2.6.18.

All results are summarized in Table \ref{tab:sumarioCaracterizacao}, which also presents resource usage average, the standard deviation between parenthesis and Naive Bayes classification in last column. Since classification is done during the benchmark and application execution, it is expected to observe some classification oscillation, which we then captured as primary and secondary workload.

\begin{table}[!h]
    \centering  
    \caption {Virtual Machine configurations.}\label{tab:configuracaoMV}
    \begin{adjustbox}{width=0.4\textwidth}
    \begin{tabu}{ccc}
        \hline\hline
        \textbf{\begin{tabular}[c]{@{}c@{}}Configuration\\ ID\end{tabular}} & \textbf{\begin{tabular}[c]{@{}c@{}}Processor\\ (VCPUs)\end{tabular}} & \textbf{\begin{tabular}[c]{@{}c@{}}Memory\\ (GB)\end{tabular}} \\ \hline
        \textit{C1}                                                         & \multirow{2}{*}{1}                                                     & 1                                                               \\  
        \textit{C2}    \vspace{4mm}                                                     &                                                                        & 2                                                               \\ 
        \textit{C3}                                                         & \multirow{2}{*}{2}                                                     & 1                                                               \\ 
        \textit{C4}                                                         &                                                                        & 2                                                               \\ \hline
    \end{tabu}
    \end{adjustbox}
\end{table}

In the SPEC characterization, running in configurations C1 and C2, with only one processor available, NB classified as CPU intensive workload, with memory or I/O fluctuations during the benchmark execution. Operations of I/O occurred when SPEC wrote statistics in control files, such as FLOPS and elapsed time. Memory classification was also expected, since the MCF has a high memory usage profile.

\begin{table}[!h]
    \centering
    \caption {Naive Bayes classification summary.}\label{tab:sumarioCaracterizacao}
    \begin{adjustbox}{width=1\textwidth}
        \begin{tabu}{cccc|cc}
            \hline\hline
            \multicolumn{4}{c}{\textbf{Average Resource Usage}}                                                                                                                                                                                                                 & \multicolumn{2}{|c}{\textbf{\begin{tabular}[c]{@{}c@{}}Naive Bayes\\ Characterization\end{tabular}}} \\ 
            \textbf{Benchmark/Application}                     & \textbf{Conf. ID} & \multicolumn{1}{c}{\textbf{\begin{tabular}[c]{@{}c@{}}CPU\\ (\%)\end{tabular}}} & \multicolumn{1}{c}{\textbf{\begin{tabular}[c]{@{}c@{}}MEM\\ (\%)\end{tabular}}} & \multicolumn{1}{|c}{\textbf{CPU (Prim/Sec)}}                 & \multicolumn{1}{c}{\textbf{MEM (Prim/Sec)}}                \\ \hline 
            \multirow{4}{*}{\textbf{SPEC}}         & \textbf{C1}       & 96 {\scriptsize($\pm$18)} & 17 {\scriptsize($\pm$5)}  & CPU+I/O & CPU+MEM                                                  \\ 
            & \textbf{C2}       & 96 {\scriptsize($\pm$18)} & 9 {\scriptsize($\pm$2)}   &  CPU+I/O & MEM                                                  \\ 
            & \textbf{C3}       & 49 {\scriptsize($\pm$12)} & 17 {\scriptsize($\pm$5)}  &  IO     & I/O+MEM                                                  \\ 
            & \textbf{C4}       & 49 {\scriptsize($\pm$11)} & 10 {\scriptsize($\pm$3)}  &   I/O+MEM & I/O                                                  \\ \hline
            \multirow{4}{*}{\textbf{LAME}}         & \textbf{C1}       & 98 {\scriptsize($\pm$15)} & 7 {\scriptsize($\pm$1)}   & CPU+I/O & I/O+CPU                                                  \\ 
            & \textbf{C2}       & 98 {\scriptsize($\pm$15)} & 5 {\scriptsize($\pm$1)}   &  CPU+I/O & I/O                                                  \\ 
            & \textbf{C3}       & 50 {\scriptsize($\pm$11)} & 7 {\scriptsize($\pm$1)}    & I/O     & I/O                                                  \\ 
            & \textbf{C4}       & 51 {\scriptsize($\pm$11)} & 5 {\scriptsize($\pm$1)}    & I/O     & I/O                                                  \\ \hline
            \multirow{4}{*}{\textbf{OpenModeller}} & \textbf{C1}       & 100 {\scriptsize($\pm$5)} & 15 {\scriptsize($\pm$1)}  &  CPU+I/O & I/O                                                  \\ 
            & \textbf{C2}       & 99 {\scriptsize($\pm$11)} & 8 {\scriptsize($\pm$1)}   & CPU+I/O & I/O                                                  \\ 
            & \textbf{C3}       & 51 {\scriptsize($\pm$11)} & 15 {\scriptsize($\pm$1)}  &  I/O+MEM & I/O                                                  \\ 
            & \textbf{C4}       & 51 {\scriptsize($\pm$11)} & 9 {\scriptsize($\pm$1)}   & I/O+MEM & I/O                                                  \\ \hline
        \end{tabu}
    \end{adjustbox}
\end{table}

\begin{figure}[!h]
    \centering
    \includegraphics[width=1\textwidth]{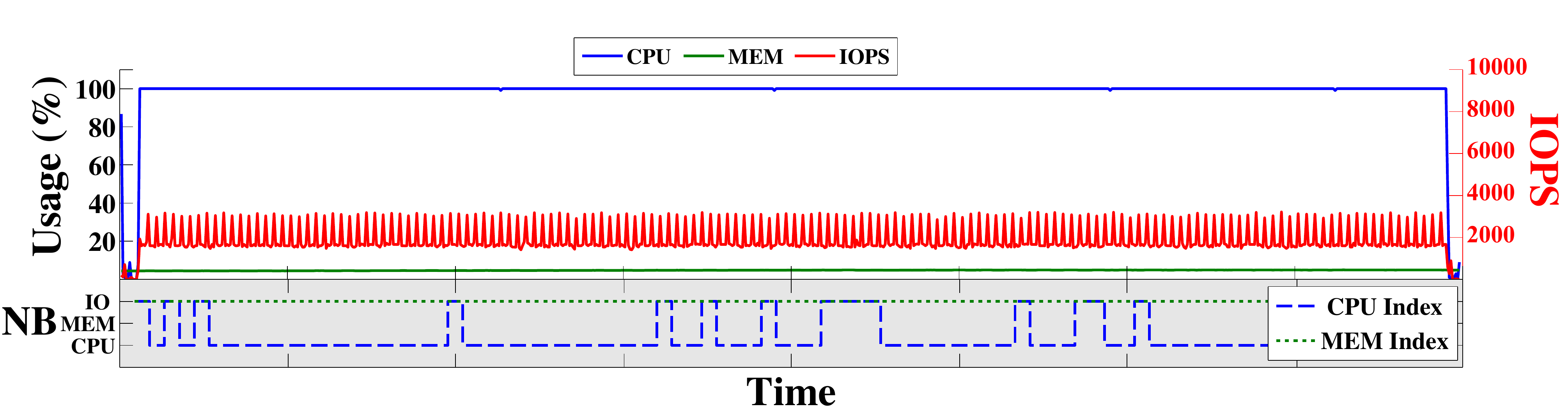}
    \caption {Characterization over time in configuration C3 running LAME.}
    \label{fig:ChacterOverTime}
\end{figure}

For the LAME benchmark, the workload profile is CPU and I/O intensive usage. While an input file is being processed, LAME creates the MP3 file, resulting in high I/O operations, with simultaneous read and write operations. Figure \ref{fig:ChacterOverTime} presents the characterization over time for configuration C3 running the workload LAME. The top of the graph contains the VM resource usage (CPU, memory, and I/O per second) and the bottom represents how NB characterized the workload at a given moment.

OpenModeller has a high CPU usage, with some memory access and I/O operations during the application initialization, when it reads the input file and during the finalization, when the benchmark writes the output file. Characterization for C1 and C2 configuration was CPU intensive, which is expected in configurations with only one processor available. The processor usage is more evident for all benchmarks/application for configurations C1 and C2 due to the availability of one processor. When adding a second processor, NB identifies other workload profiles. 

The NB's asymptotic complexity is linear which, as previously discussed, is necessary for any characterization strategy in cloud computing. Considering the discretization steps and the probability computation, the complexity is $\Theta(n+k)$, where $k$ is the number of indexes to be discretized and $n$ is the number of classes to be evaluated.

By using the NB characterization we identify when a VM can be moved across physical hosts. Using NB characterization, we can identify the primary workload and, instead of usual classification as CPU, MEM, I/O or IDLE, it is classified as suitable to LM or non-suitable to LM (NLM).

\subsection{Orchestration Analysis}

This experiment evaluates the orchestration considering suitable moments to trigger live migration. A live migration representation graph illustrates the workload behavior over time and the moments when live consolidation were submitted and the instant when live migration actually occurred.

\subsubsection{Benchmark Experiments} \label{sec:BenchEval}

\begin{figure*}[t!]
    \centering
    \includegraphics[width=\textwidth]{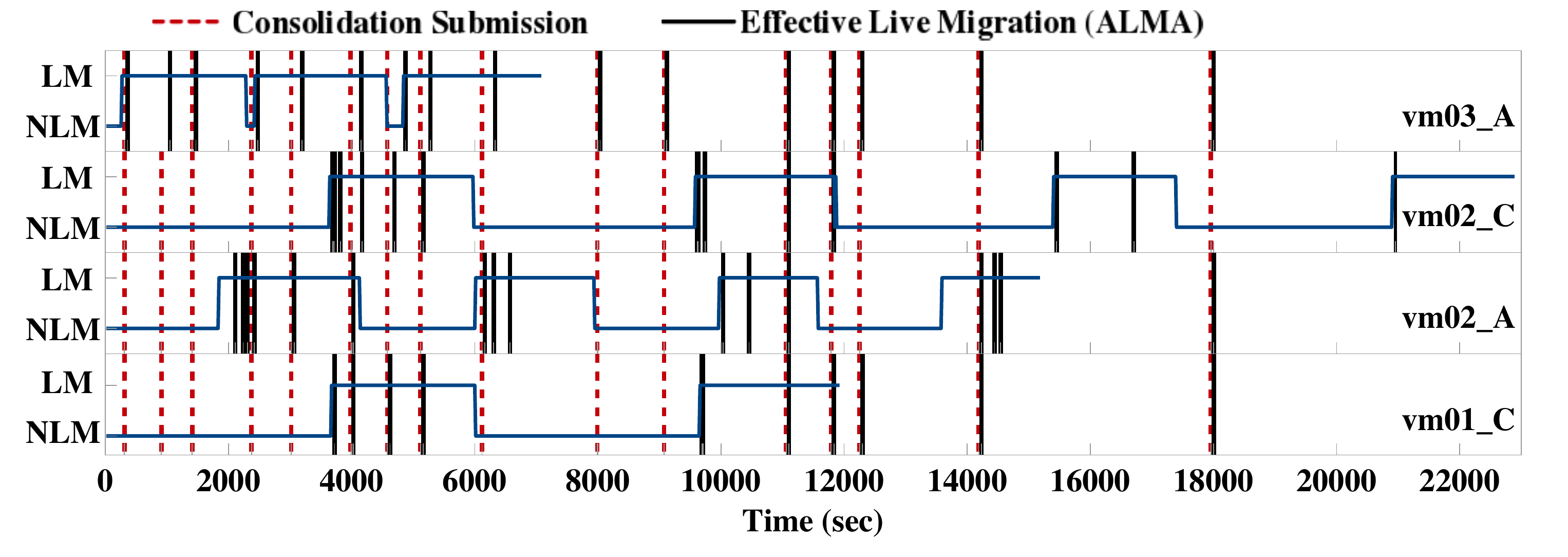}
    \caption{Cycle accuracy identification diagram for benchmarks.}
    \label{fig:LMALMA_40Bench}
\end{figure*}

Figure \ref{fig:LMALMA_40Bench} presents the migration diagram for the four VMs running benchmarks. Line in blue is the workload behavior over time, where valleys are periods not suitable to trigger live migration (NLM) and peaks are periods where the workloads are suitable to live migration (LM). The workload is executed at the same time across all VMs. Dashed lines in red represent consolidation instants. Lines in black are instants where ALMA actually triggered live migration. 

In order to compare the consolidation strategy under control of our architecture and without any surveillance we run two sets of experiments. In the first set, VMs were actually consolidated in instants represented in dashed red lines and we left the workload run to the end. During the second set, our architecture was in place and according to the workload and the cycle analysis, it triggered or postponed live migration to a better instant (represented by the black lines in the figure).

\begin{table}[!t]
\centering
\caption {Results with four VMs running benchmarks.}
\begin{adjustbox}{width=0.75\textwidth}
\begin{tabu}{ccccc}
\hline\hline
\textbf{Metric}                                                                              & \textbf{\begin{tabular}[c]{@{}c@{}}Virtual\\ Machine\end{tabular}} & \textbf{\begin{tabular}[c]{@{}c@{}}Traditional\\ Consolidation\end{tabular}} & \textbf{ALMA} & \textbf{\begin{tabular}[c]{@{}c@{}}Reduction \\ (\%)\end{tabular}} \\ \hline 
\multirow{4}{*}{\textbf{\begin{tabular}[c]{@{}c@{}}Downtime\\ (sec)\end{tabular}}}            & \textbf{vm03\_A}                                                   & 20.06                                                                       & 20.44         & -1.87                                                            \\  
                                         & \textbf{vm02\_C}& 18.63                                                                       & 17.75         & 4.70                                                             \\  
                                         & \textbf{vm02\_A}                                                   & 20.75                                                                       & 23.69         & -14.16                                                           \\  
                                          & \textbf{vm01\_C}                                                   & 19.25                                                                       & 18.94         & 1.62                                                             \vspace{4mm}\\ 
\multirow{4}{*}{\textbf{\begin{tabular}[c]{@{}c@{}}Live Migration\\ Time\\ (sec)\end{tabular}}} & \textbf{vm03\_A}                                                   & 28.81                                                                       & 12.00         & 58.35                                                            \\  
                                                     & \textbf{vm02\_C}                                                   & 87.56                                                                       & 42.31         & 51.68                                                            \\  
                                                      & \textbf{vm02\_A}                                                   & 43.81                                                                       & 11.13         & 74.61                                                            \\ 
                                                      & \textbf{vm01\_C}                                                   & 54.31                                                                       & 26.81         & 50.63                                                            \vspace{4mm}\\ 
\multicolumn{2}{c}{\textbf{\begin{tabular}[c]{@{}c@{}}Data \\ Traffic\\ (MB)\end{tabular}}}                                                                  & 11,557.50                                                                    & 9,159.60       & 21.56                                                            \\ \hline
\end{tabu}
\end{adjustbox}
\label{tab:ComparacaoMetricas40Bench}
\end{table}

Ideally, when ALMA is in place, live migration (lines in black) should be triggered during the peaks. In this experiment, our architecture was able to migrate VMs at suitable workload moments, thus reducing data transferred and live migration time (Table \ref{tab:ComparacaoMetricas40Bench}). The reduction in live migration time was up to 74\% (vm02\_A) and data traffic reduction was up to 21\%, representing a reduction of about 2.3 GB. For the downtime metric, with 95\% of confidence, it is not possible to infer any improvements when using ALMA or not.  

\subsubsection{Application Experiments} \label{sec:ApplEval}

We used two scientific applications, BRAMS and OpenModeller running in vm02\_C and vm03\_A, respectively  and a typical cloud workload, represented by Hadoop cluster, running in vm01\_B, vm02\_C and vm01\_C. The vm01\_B VM was not moved from the physical host because it already was in one of the physical hosts in which the workload was consolidated. That is the reason why metrics of this VM were suppressed from presented results.

\begin{figure*}[ht!]
    \centering
    \includegraphics[width=\textwidth]{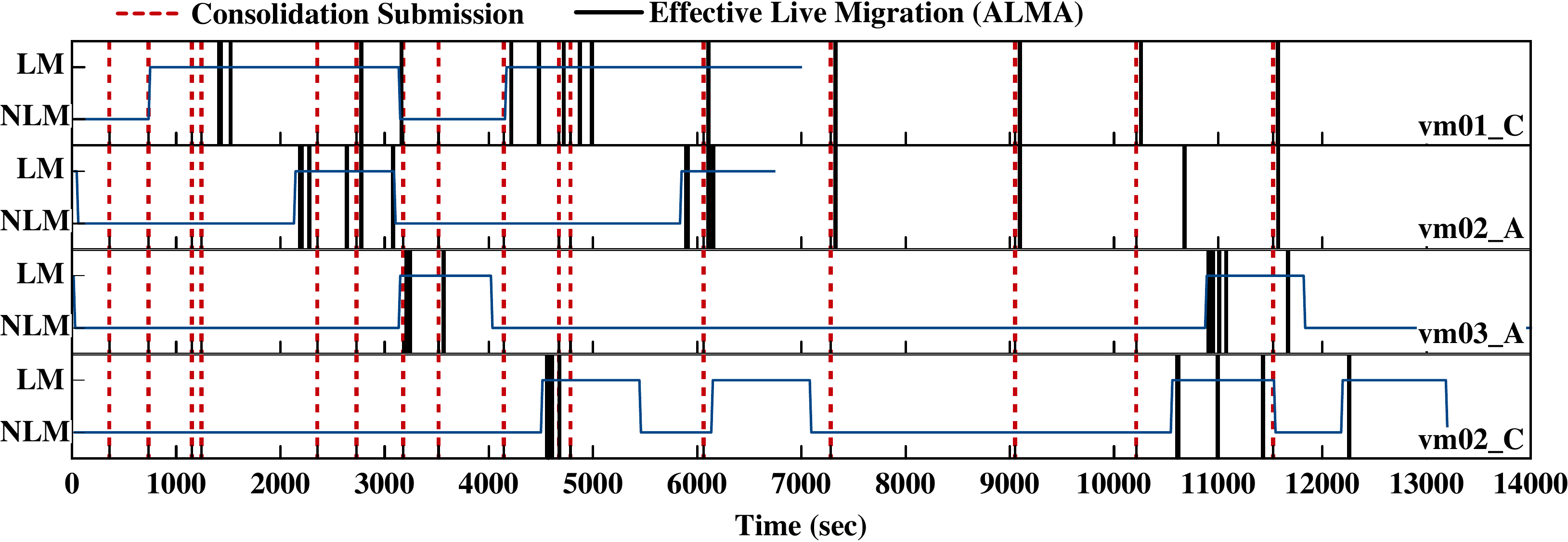}
    \caption{Cycle accuracy identification diagram for applications.}
    \label{fig:LMALMA_40Aplic}
\end{figure*}

Figure \ref{fig:LMALMA_40Aplic} shows long suitable periods to live migration, such as the one of the vm01\_C. There are also workloads with long periods not suitable to live migration, like vm03\_A's workload, and complex cycles as observed in vm02\_C. However, even in this scenario, ALMA was able to identify and successfully postpone the live migration to a suitable moment.

The ALMA accuracy leads to the results presented in Table \ref{tab:ComparacaoMetricas40Aplic}. Reduction in live migration time (up to 67\%) and amount of data transferred in network (up to 62\%) were significant. This result is due to the Hadoop cluster behavior, which exchanges larges amount of data across the cluster nodes. This application, particularly, was benefited by our architecture, as observed in Table \ref{tab:ComparacaoMetricas40Aplic}. 

\begin{table}[!h]
\centering
\caption {Results with four VMs running real applications.}
\begin{adjustbox}{width=0.75\textwidth}
\begin{tabu}{ccccc}
\hline\hline
\textbf{Metric} & \textbf{\begin{tabular}[c]{@{}c@{}}Virtual\\ Machine\end{tabular}} & \textbf{\begin{tabular}[c]{@{}c@{}}Traditional\\ Consolidation\end{tabular}} & \textbf{ALMA} & \textbf{\begin{tabular}[c]{@{}c@{}}Reduction \\ (\%)\end{tabular}} \\ \hline 
\multirow{4}{*}{\textbf{\begin{tabular}[c]{@{}c@{}}Downtime\\ (sec)\end{tabular}}}            & \textbf{vm03\_A (OpenModeller)}                                                   & 21.80                                                                       & 23.00         & -5.50                                                            \\ 
                               & \textbf{vm02\_C (BRAMS)}                                                   & 22.60                                                                       & 20.73         & 8.26                                                             \\ 
                               & \textbf{vm01\_C (Hadoop)}                                                   & 19.07                                                                       & 22.33         & -17.13                                                           \\ 
                               & \textbf{vm02\_A (Hadoop)}                                                   & 12.67                                                                       & 17.20         & -35.79\vspace{4mm}                                                          \\ 
\multirow{4}{*}{\textbf{\begin{tabular}[c]{@{}c@{}}Live Migration\\ Time\\ (sec)\end{tabular}}} & \textbf{vm03\_A (OpenModeller)}                                                   & 31.27                                                                       & 12.73         & 59.28                                                            \\ 
                                 & \textbf{vm02\_C (BRAMS)}                                                   & 12.93                                                                       & 10.60         & 18.04                                                            \\  
                                  & \textbf{vm01\_C (Hadoop)}                                                   & 39.20                                                                       & 18.67         & 52.38                                                            \\ 
                                  & \textbf{vm02\_A (Hadoop)}                                                   & 38.20                                                                       & 12.40         & 67.54                                                            \vspace{4mm}\\ 
\multicolumn{2}{c}{\textbf{\begin{tabular}[c]{@{}c@{}}Data \\ Traffic\\ (MB)\end{tabular}}}                                                                  & 14,566.47                                                                    & 5,504.98       & 62.21                                                            \\ \hline
\end{tabu}
\end{adjustbox}

\label{tab:ComparacaoMetricas40Aplic}
\end{table}

The application behavior is not known {\it a priori} (as opposed to the benchmarks evaluation) and it is sensitive to the initial setup, such as command line parameters and the input data. In both experiments, downtime did not show improvements or deterioration. Statistically, with 95\% of confidence, it is not possible to determine if using ALMA or not can achieve performance improvements. 

The explanation for this is in the TCP behavior and how the hypervisor contacts the network devices that a given IP address is hosted by a given physical host. Once a VM is moved across physical hosts, the hypervisor needs to update the ARP table and, to this end, it sends an ICMP packet to the network gateway. This process is not part of the migration algorithm; it is an independent process. Moreover, downtime is sensitive to TCP. Our results corroborate observations from Kikuchi and Matsumoto \cite{Kikuchi:2012}: when a packet does not arrive to the destiny, sender will try again when retransmission time out (RTO) ends. The RTO is computed using round time trip (RTT) which is, initially, equal to three seconds. Every time a retransmission is needed, RTO value is doubled, increasing downtime.

\subsection{Scalability} \label{sec:Scalability}

Scalability is a factor to be considered when dealing with cloud computing environments. Among the main features of such environment we can cite resource elasticity, which enables the user to increase or decrease the amount of available computational resources in response to a given demand \cite{Armbrust:2009}. Also, users could add or remove VMs to their environment. These features make cloud computing environments sensitive to solutions with low scalability characteristic.

\begin{figure}[!h]
    \centering
    \includegraphics[width=0.7\textwidth]{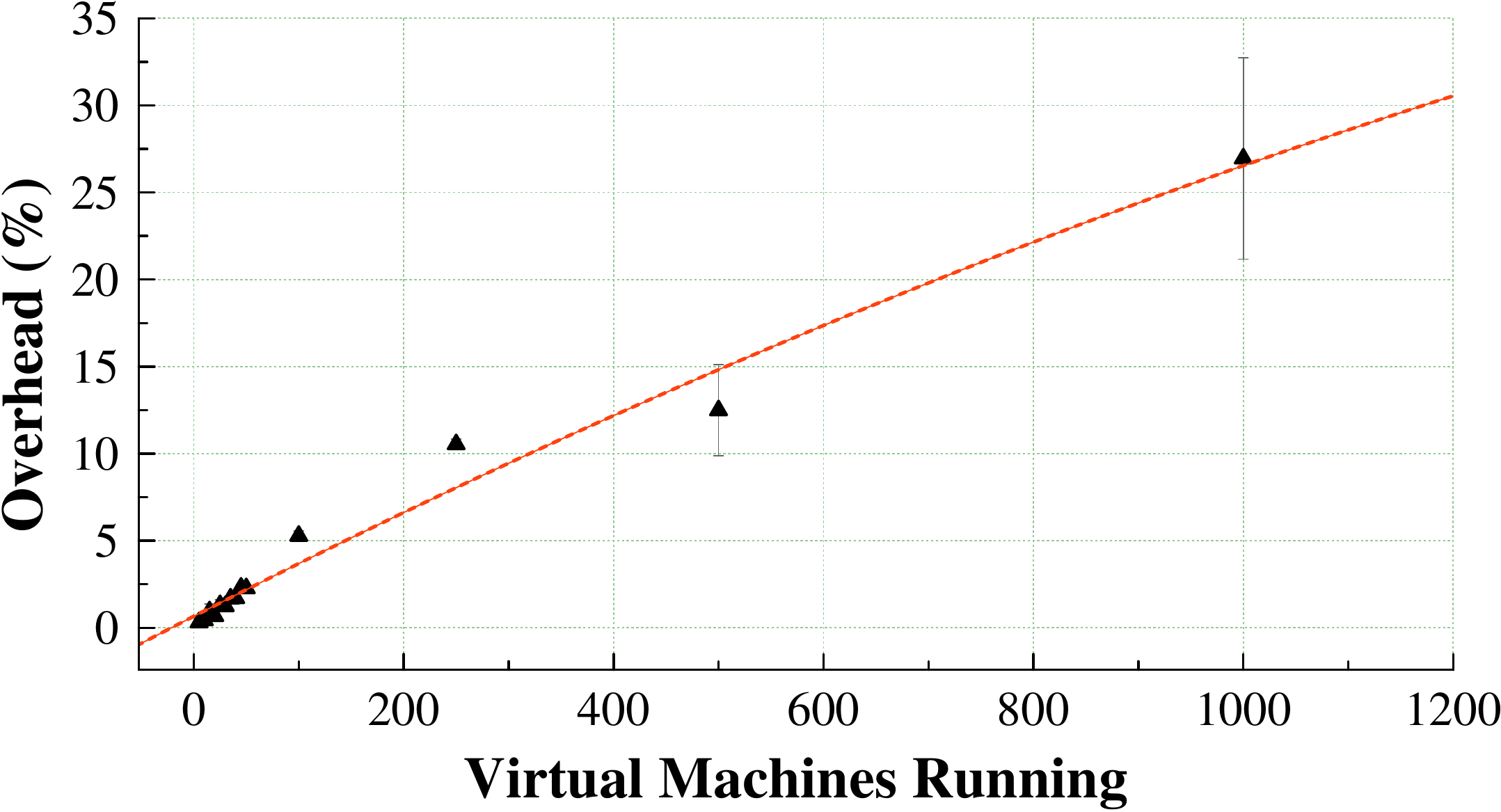}
    \caption{Overhead caused by LMCM with data from up to thousand VMs.}
    \label{fig:SobrecargaMCLM}
\end{figure}

In this section we present an analysis of scalability of our solution. Despite the fact that previous experiments were run in a real private cloud, performing a scalability test would require a large testbed, which we do not have access to. Therefore, we used traces of the previous executed benchmark experiments to measure the overhead caused by ALMA, specifically the Live Migration Control Module (LMCM) that performs the classification and cycle analysis.

The baseline used to infer the overhead is the Linux kernel compilation with no other processes competing for resources. Next, traces were submitted to the LMCM and the amount of extra time to perform the same compilation was considered as the overhead. For each VM a new process was created inside LMCM and we started the analysis with five VMs and gradually increased to up to thousand VMs. Figure \ref{fig:SobrecargaMCLM} shows the results from ten-run experiments which highlight the fact that the overhead has a linear tendency (line in red) that increases proportionally with the increase of VMs. The overhead has increased, in average, 0.21\% for every five VMs added. According to this result, and in this configuration testbed, LMCM saturation would be achieved with 1,800 VMs running and submitting the live migration request at the same time. We considered the saturation when the overhead caused by LMCM reached 100\% when compiling the linux kernel.

\subsubsection{Data Gathering Overhead in Virtual Machines} \label{sec:VMOverhead}
The overhead imposed by index data gathering in VMs needs to be considered. As mentioned earlier, data indexes were collected by SNMP version 2. For each SNMP request, a script is executed, which returns a given value according to the index requested. The overhead evaluation conducted in this section is similar to the evaluation that has been conducted to infer the LMCM overhead. We changed the VM configuration during the experiment (processor and memory) and our main objective is to observe and quantify the overhead caused by index data gathering according to the amount of available computational resource.

\begin{figure}[!h]
    \centering
    \begin{subfigure}{0.48\textwidth}
        \includegraphics[width=\textwidth]{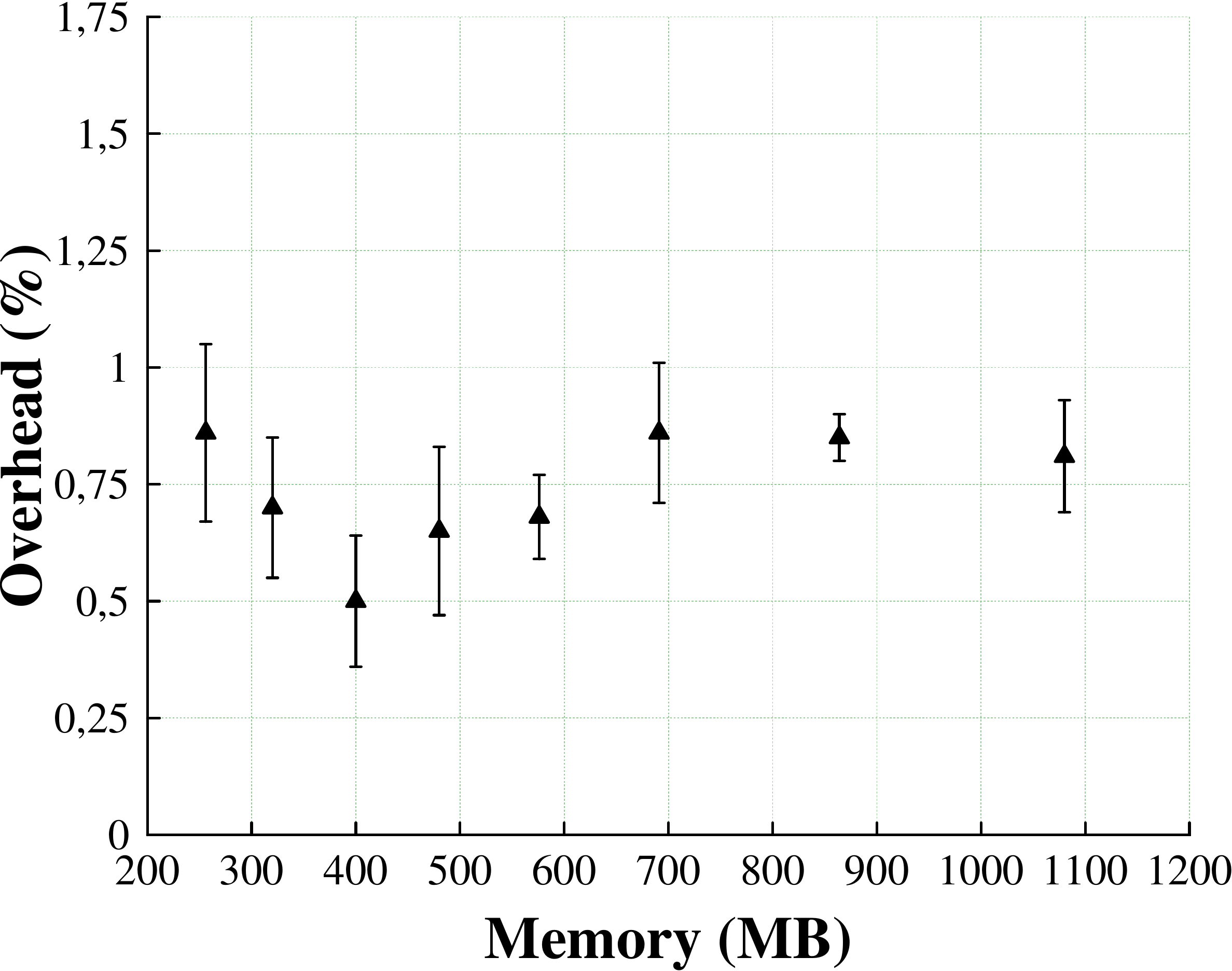}
        \label{fig:sobrecarga1VCP}
        \caption{One processor.}    
    \end{subfigure}
    ~
    \begin{subfigure}{0.48\textwidth}
        \includegraphics[width=\textwidth]{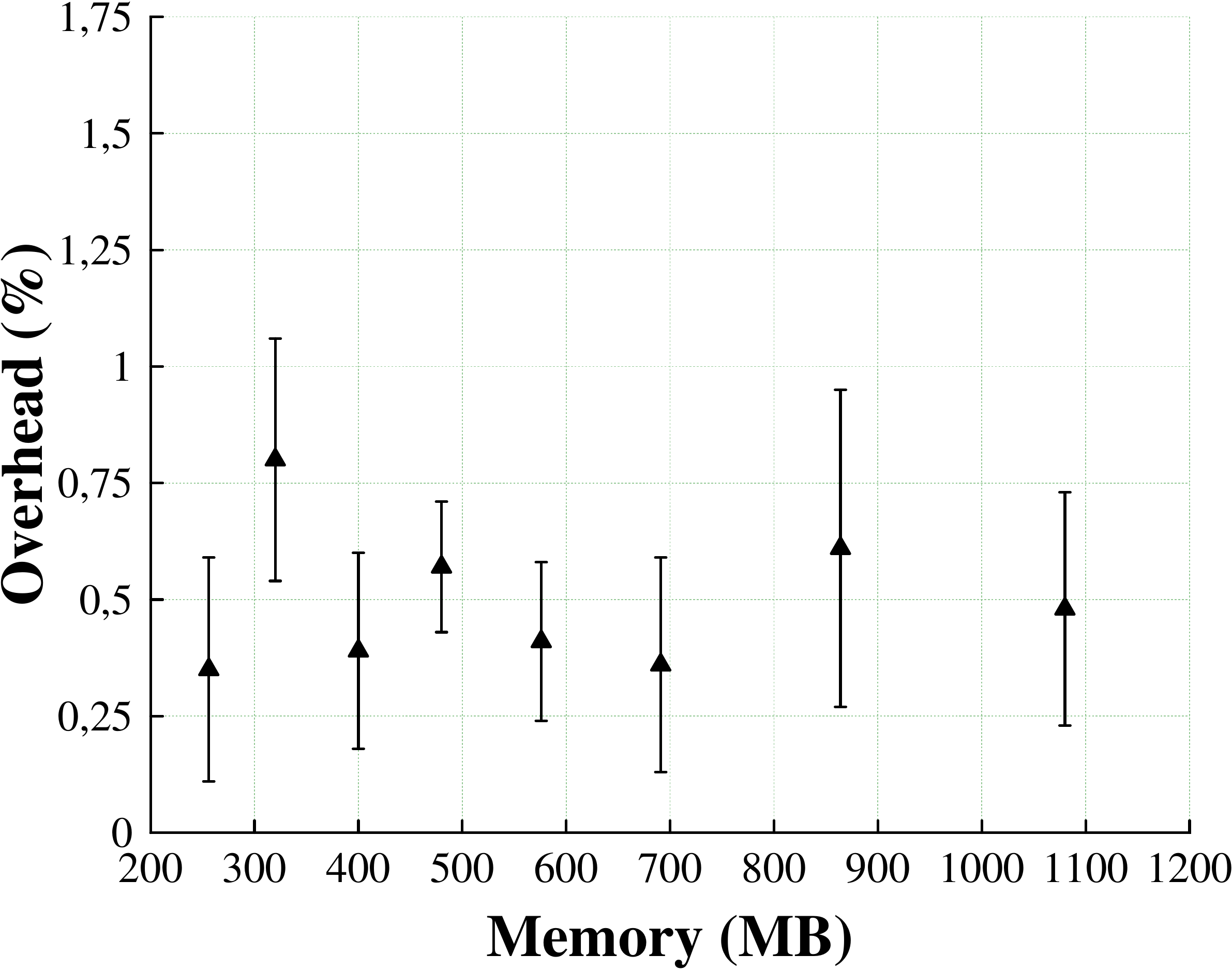}
        \label{fig:sobrecarga2VCP}
        \caption{Two processors.}        
    \end{subfigure}
    \caption{Overhead in virtual machine.}
    \label{fig:SobrecargaMVs}
\end{figure}

The first experiment set was conducted using one VM with one processor and memory increased from 256 MB to 1,080 MB. Results are presented in Figure \ref{fig:SobrecargaMVs} and indicate the overhead is about 0.75\% and 0.5\% for VMs with one and two processors, respectively. The overhead is constant with small fluctuation when varying the memory size. However, it is more sensitive to the available processors.

\section{Related Work} \label{sec:Sec05}

The main research efforts on live migration overhead for cloud environments are related to VM workload consolidation techniques \cite{ye2015profiling,ye2011live,Xu:2013}. The motivation comes from the substantial cost reduction for service providers that can be achieved using optimized consolidation techniques. Due to the different aspects of existing efforts in this area, we split them into two categories. The first one is on live migration overhead as a factor to trigger live migration. The efforts in second category present strategies to control the live migration to avoid network congestion caused by massive migrations.

\subsection{Live Migration with Overhead Constraints}
As discussed earlier, consolidation deals with the selection and transfer of VMs to a common physical host in order to reduce power consumption of the source physical hosts. Since this is an NP-hard problem, several solutions based on heuristics are available in literature. Resource reservation was also considered as a mechanism to optimize migrations \cite{ye2011live}. However, most of solutions do not consider live migration overhead in infrastructure and VMs collocated.

Xu et al. \cite{Xu:2013} present a strategy that takes into account the computational cost of live migration in physical hosts (source and target) that are involved in the migration process and in collocated VMs. This strategy, called iAware, aims to avoid service level agreements violation, preventing that a given live migration harms other VMs. Authors also present a strategy to model the impact of live migration in collocated VMs, which is based on available physical resources and the amount of interruptions generated by VMs. Similar to our work, iAware can be embedded into existing consolidation or load balance strategies. Furthermore, both strategies aim to reduce performance degradation caused by live migration.

Despite the features in common, iAware does not consider the VM workload as it only relies on resource usage of the physical host. Live migration itself has a computational cost, therefore, postponing it to more suitable moment according to the VM workload can reduce overhead in physical host.

Verma et al. \cite{Verma:2011} present CosMig, which is a model for live migrations, including time estimation to perform them. CosMig is based on processor and memory usage parameters and determines the live migration impact of a VM. Verma et al. also identified that: (1) an effective live migration model must take into account application behavior, (2) only live migration does not improve application performance; other factors can promote performance improvement such as target host computing power and VM memory fragmentation. The main similarity of CosMig and our work is the evaluation of live migration in VM workload. Despite the fact that metrics used to model live migration impact are different, both studies present models to infer live migration impact in workload. A fundamental difference of is how information about live migration impact is used by the proposed strategy. In CosMig, the question asked is related to \textit{``if''} live migration of a given VM will lead to performance gains or not. On the other hand, ALMA asks \textit{``when''} a live migration can be performed in order to avoid infrastructure damage and, consequently, in application.

Finally, Stage and Setzer \cite{Stage:2009} introduce a live migration scheduling strategy that classifies migrations according to the current workload and identifies the minimal network resources to perform a migration. According to the authors, a migration of a single VM can consume significant network bandwidth during a long period (about 500 Mb/s for ten seconds to migrate a VM running a web server). The architecture presented by Stage and Setzer has similarities with our work. Like in ALMA, there is a live migration scheduler management, which decides when a VM can be migrated. However, their architecture only observes network parameters (available bandwidth and a live migration time constraint). Also, there is a workload classifier based on the following attributes: (i) predictable:  workload is considered predictable when its behavior has a reliable prognosis for a given period; (ii) tendentious: refers to fluctuations of a tendency; and (iii) cyclic: indicates how often a pattern occurs in a given workload. The main difference from our work is that, in Stage and Setzer work, live migration will take place according to the network bandwidth consumption. From the estimative based on workload type and live migration duration threshold, which can be defined by user or service provider, the architecture schedules live migrations in order to meet live migration duration threshold. In addition, the characterization of Stage and Setzer aims to group VMs with similar workload and performs the live migration of these groups. In ALMA, the workload characterization is the main criterion to define the suitable moment to trigger live migrations. 

\subsection{Live Migration Control Strategies}
Beloglazov and Buyya \cite{Beloglazov:2013} propose a dynamic strategy for VM consolidation that considers suitable moments to perform live migrations. Their goal is to minimize power consumption and maximize quality of service (QoS) delivered by service provider which, according to the authors, composes the trade-off between energy and performance. Their strategy identifies physical hosts overloaded and live migrations intervals are defined in order to keep QoS.
In our work, ALMA can postpone a live migration according to the VM workload and in Beloglazov and Buyya study, live migrations can be postponed according to the physical host workload.

Ye et al. \cite{Ye:2012} present a framework, called VC-Migration, which controls live migrations in a cluster composed of VMs. The VC-Migration has strategies previously configured which decides how many VMs (granularity) will be considered for migration in a given moment. The decision is based on current computational resource usage of physical hosts. The strategies defined by the framework are:
\begin{itemize}
    \item \textbf{Concurrent migration:} this strategy performs the live migration of several VMs, simultaneously, running in the same cluster;
    \item \textbf{Mutual migration:} strategy which is applied when physical hosts involved in live migration process have VMs moved between each other; 
    \item \textbf{Homogeneous migration in multi-cluster:} strategy applied when several virtual clusters, with the same number of VMs, are being migrated;
    \item \textbf{Heterogeneous migration in multi-cluster:} same strategy as homogeneous migration, but virtual clusters have different sizes.
\end{itemize}

The framework chooses the best strategy according to the number of VMs being migrated and network bandwidth consumption. Authors argue that application interdependence, which is common in a cluster environment, reduces the infrastructure impact for network and applications.
\section{Concluding Remarks} \label{sec:Sec06}

Live migration algorithms are known to be sensitive to memory usage. However, during an application execution these algorithms can present periods of high memory usage or high processor usage. These periods can float according to the day of the week, period of the year, or even with application input. Therefore, the challenge is to identify workloads with cyclic pattern and, once the cycle is identified, how to postpone live migrations to reduce their overhead. 

We proposed and evaluated a migration strategy and architecture using a private cloud running benchmarks and real applications. The architecture was able to reduce up to 74\% and 62\% in live migration time and data traffic, respectively. The scalability analysis showed a host with 6 GB of memory is capable of handling data of up to 1,800 VMs. Based on evaluation results, our main finding is that using workload cycle recognition it is possible to choose suitable moments to trigger live migration, thus leading to a significant reduction in migration time and data traffic, confirming our hypothesis.

\bibliographystyle{elsarticle-num} 
\bibliography{references}




\end{document}